\documentclass[%
 reprint,
 amsmath,amssymb,
 aps,
]{revtex4-2}
 
\usepackage{graphicx}% Include figure files
\usepackage{dcolumn}% Align table columns on decimal point
\usepackage{bm}% bold math
\usepackage{natbib}
\setcitestyle{super,sort,compress,comma}

\usepackage[colorlinks]{hyperref}
\usepackage[nameinlink]{cleveref} 
\usepackage{titlesec}
\usepackage{booktabs,tabularx}
\usepackage{enumitem}
\usepackage{url}
\usepackage{microtype}
\usepackage[english=usenglishmax]{hyphsubst}
\usepackage{subfigure}
\usepackage[left]{lineno}
\usepackage{xcolor}
\usepackage{listings}
\usepackage{verbatim}
\usepackage[T1]{fontenc}
\usepackage{enumitem}

\makeatletter
\renewcommand{\fnum@figure}{\textbf{\figurename~\thefigure}}
\def\@caption@fignum@sep{ }
\makeatother

\makeatletter
\renewcommand{\fnum@table}{\textbf{\tablename~\thetable}}
\makeatother

\renewcommand{\figurename}{\textbf{Fig.}}
\renewcommand{\tablename}{Tab.}

\setlist[enumerate]{itemsep=0.5em, topsep=0.5em, partopsep=0em, parsep=0em}

\hypersetup{
    colorlinks=true,
    linkcolor=blue,     
    citecolor=blue, 
    urlcolor=blue   
}

\titleformat*{\subsubsection}{\scriptsize\bfseries}

\titlespacing*{\section}{0pt}{0.1\baselineskip}{0.2\baselineskip}

\titlespacing*{\section}
{0pt}{2.5ex}{2.5ex}
\titlespacing*{\subsection}
{0pt}{2ex}{2ex}
\titlespacing*{\subsubsection}
{0pt}{2ex}{2ex}

\begin{document}

\title{A unified moment tensor potential for silicon, oxygen, and silica}

\author{Karim Zongo$^{1, *}$}
\author{Hao Sun$^{2}$}
\author{Claudiane Ouellet-Plamondon$^{1}$}
\author{Laurent Karim Béland$^{2, *}$} 

\affiliation{ \vspace{10pt} \textnormal{$^{1}$Département de génie de la construction, École de Technologie Supérieure, Montréal, QC, Canada  \\ $^{2}$Department of Mechanical and Materials Engineering, Queen's university, Kingston, ON, Canada  \vspace{10pt} \\ Correspondence: Karim Zongo (karim.zongo.2@ens.etsmtl.ca) or Laurent Karim Béland (laurent.beland@queensu.ca) }}

\begin{abstract}
\section*{Abstract}
Si and its oxides have been extensively explored in theoretical research due to their technological importance. Simultaneously describing interatomic interactions within both Si and SiO$_2$ without the use of \textit{ab initio} methods is considered challenging, given the charge transfers involved. Herein, this challenge is overcome by developing a unified machine learning interatomic potentials describing the Si/ SiO$_2$/ O system, based on the moment tensor potential (MTP) framework. This MTP is trained using a comprehensive database generated using density functional theory simulations, encompassing diverse crystal structures, point defects, extended defects, and disordered structure. Extensive testing of the MTP is performed, indicating it can describe static and dynamic features of very diverse Si, O, and SiO$_2$ atomic structures with a degree of fidelity approaching that of DFT.
\end{abstract}

\maketitle
\section*{Introduction}
Si/SiO$_2$ interfaces are ubiquitous in semiconductor manufacturing, which includes metal–oxide–semiconductor field-effect transistors \cite{arns1998other}, nanowire- and nanodot-transistors. The formation of SiO$_2$ layers involves charge transfer during the oxidation of Si substrates \cite{bauza2001thermal}. Additionally, siliceous materials—including clay minerals and cement, which comprise Si, O, and SiO$_2$ components—are governed by interactions that entail similar charge transfer. Abundant past theoretical research has focused on understanding Si, its oxides, the formation of SiO$_2$ multilayered structures, early oxidation rates, and amorphization of oxide layers \cite{pasquarello1998interface,ganster2010molecular,ganster2012first,cvitkovich2023dynamic,cvitkovich2022computational,salles2017strain,takahashi2014molecular}. These studies predominantly relied on electronic density functional theory (DFT) \cite{kohn1965self, sholl2022density, goedecker1999linear} and nonflexible classical potentials \cite{torrens2012interatomic,pizzagalli2004classical}. However, simulating large systems with multiple components, charge transfer, and hetero-interfacial systems poses challenges within these frameworks. An ideal modeling approach should explicitly or implicitly capture charge transfer without compromising accuracy or incurring prohibitively large computational costs. Existing charge equilibration potentials like ReaxFF \cite{van2001reaxff, van2003reaxffsio} and COMB \cite{phillpot2018charge,yu2007charge, liang2012variable}, while being capable of describing chemical interactions during MD simulations, tend to have a limited ability to describe mechanical properties of materials \cite{van2003reaxffsio,yu2007charge, shan2010second} unless special reparametrization is applied.

Recent advances, such as linear-scaling DFT \cite{goedecker1999linear, hine2009linear} and machine learning (ML) force fields--e.g. the Gaussian approximation potentials (GAP) \cite{bartok2010gaussian} and artiﬁcial neural networks \cite{behler2007generalized,behler2021four}-- lift the limitations of traditional methods. ML force fields have demonstrated high accuracy in modeling Si \cite{bartok2018machine,li2020unified,hu2019accurate,babaei2019machine}  and many other elements \cite{dragoni2018achieving,szlachta2014accuracy,rowe2020accurate,smith2021automated,deringer2020general}. Similar progress has been made in improving interatomic interaction descriptions in Si oxides \cite{novikov2019improving,erhard2022machine,li2018comparison,balyakin2020deep} and metal oxides \cite{artrith2011high, artrith2016implementation,kandy2023comparing,kobayashi2022machine}. However, jointly describing compounds and their constituents using ML force fields presents challenges due to the disjointed configurational space of multi-phase forms and the need to handle charge transfer.
ML force fields \cite{behler2016perspective,das2019machine,mishin2021machine,onat2020sensitivity,friederich2021machine} combine a descriptor to a regression procedure to encode geometry and \textit{ab initio} properties, usually omitting explicit electronic structures. A previous study focusing on modeling SiO$_2$ using the moment tensor potential (MTP) suggests incorporating additional reference data is preferable to adding explicit charge equilibration for long-range interactions \cite{novikov2019improving}.

The novelty of this article is a MTP \cite{shapeev2016moment,novikov2020mlip} that jointly describes interatomic interactions in SiO$_2$ and its constituents (Si and O), enabling the representation of multiple charge states. The developed MTP for Si/ O/ SiO$_2$ systems is parameterized using an \textit{ab initio} database containing diverse crystal structures, point defects, extended defects, and disordered structures. This MTP is then utilized for molecular statics (MS) and molecular dynamics (MD) simulations to investigate crystalline, interfaces, amorphous, and liquid states of Si and SiO$_2$. These test simulations indicate that the MTP can provide a unified description of these disjoint systems.

%\newpage
\section*{Results}
Our analysis encompasses both MS and MD simulations. MS results include cohesive energy, lattice constant, elastic constant, defect formation energies, interface relaxation, as well as linear and planar defect. The MD results bracket vacancy diffusion coefficients, melting point,interfaces, as well as the liquid and amorphous structures of Si and SiO$_2$. The outcome of these runs are compared against the reference method (DFT) and those derived from the semi-classical potential.
\subsection*{Cohesive energy, elastic constant, point defects and extended defects}
First, the state equations of various Si and SiO$_2$ polymorphs are presented as cohesive energy vs lattice parameter. Please see the methodology section for calculation details. Additionally, the cohesive energy values for molecular oxygen, both O$_2$ and O$_3$, are provided. As shown in Fig. \textcolor{blue}{1}, the MTP replicated the cohesive energies of the references states with remarkable accuracy. The Tab. \ref{tab:L_param} compares lattice parameters predicted by the MTP, COMB, and ReaxFF models with benchmark and experimental data. Remarkably, the MTP predictions show excellent agreement with both the benchmark and experimental data.
\indent

%\begin{comment}
\begin{figure}
\vspace*{-1cm}
    %\includegraphics[width=0.32\textwidth]{EOSALL.png}
    %\centering
    \includegraphics[width=6.8cm]{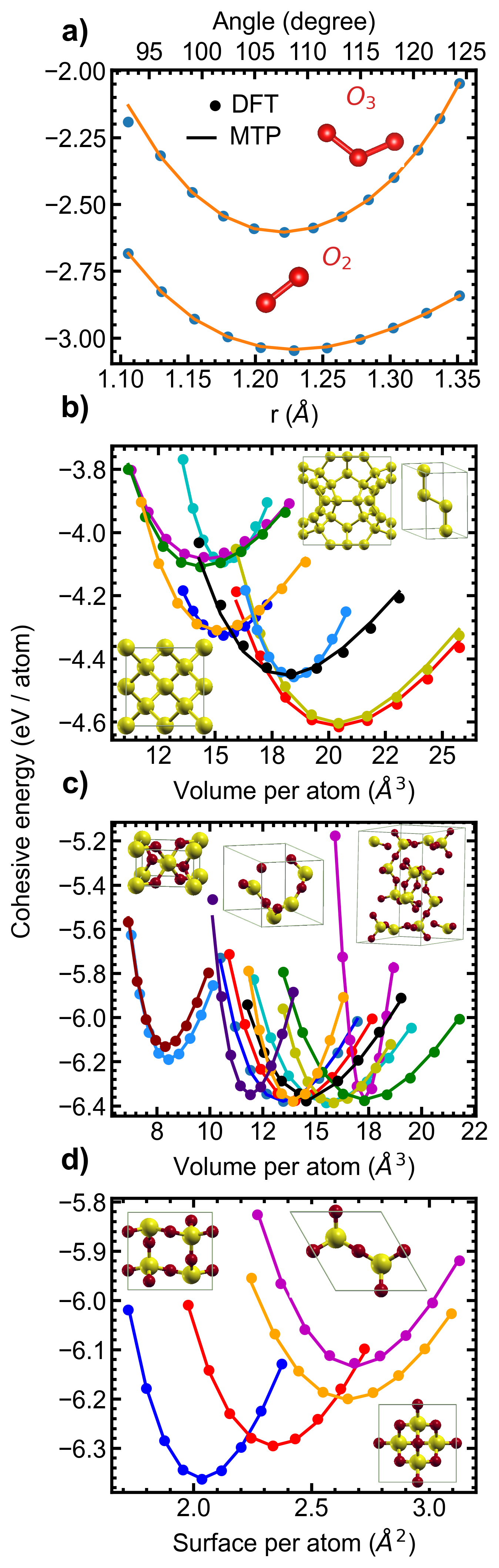}
    \caption{\label{fig:EOS} \textbf{Bond and angle energy of oxygen molecules, and equation of state of silicon and silica polymorphs}. \textbf{a} Bond and angle energy of dioxygen and ozone as calculated using MTP (lines) and DFT (dots). \textbf{b-d} Energy-volume relationships in crystalline SiO$_2$ and Si polymorphs calculated using the MTP (lines) and DFT (dots). 3D Si polymorphs \textbf{b}, 3D \textbf{c} and 2D \textbf{d} SiO$_2$ structures were considered. Details pertaining to these crystal structures are available as S.I. Agreement between MTP and DFT is excellent. Insets: illustration of selected polymorphs' crystal structures.}  
\end{figure}
%\end{comment}

\begin{table} 
\centering
\begin{tabular}{lcccccc}

\multicolumn{6}{c}{ } \\ 

 & & Exp & DFT & MTP & COMB & ReaxFF \\
\hline
$\alpha$ Si & a & 5.4300\cite{brandes2013smithells} & 5.4690 & 5.4682 & 5.4399 & 5.4635\\ 
$\beta$-Si & a & 4.665\cite{mcmahon1994pressure} & 4.8090 & 4.8170 & 4.8242 &  4.7650  \\
                & c &2.565 & 2.6554 & 2.6493 & 2.6533 &  2.6207 \\
HD & a & &  3.8508 & 3.8601  & 3.8411 &  3.8635  \\
                & c &  & 6.2706 & 6.3028 & 6.2718 &  6.3084  \\
HCP & a & &  2.6961 & 2.8038  & 2.8065 &  2.7192  \\
                & c &  & 4.5577 & 4.6013 & 4.6057 &  4.4624  \\
SH & a & 2.5530\cite{mcmahon1994pressure} & 2.6458 & 2.6478  & 2.7232 &  2.9162  \\ 
                & c & 2.3820 & 2.4840 & 2.4861 & 2.5569 &  2.7382  \\
ST12 & a &  & 5.6763 & 5.6639 & 5.7680 &  5.65236  \\
                & c &  & 6.8308 & 6.7607 & 6.8849 &  6.7469  \\
BC8 & a &  & 5.7657 & 5.7662 & 5.7539 &  5.6405  \\
                & c &  & 4.7077 & 4.7081 & 4.6980 &  4.6054  \\
C24 & a & 3.82\cite{kim2015synthesis} & 3.8510 & 3.8533 & 3.8482 &  3.7800  \\
                & b & 10.70 & 10.2298 & 10.7874 & 10.7732 &  10.5821  \\
                & c & 12.63 & 12.7483 & 12.7281 & 12.7113 &  12.4859  \\
C46 & a & 10.3550\cite{adams1994wide} & 10.2298 & 10.2777 & 10.1513 & 10.1761\\
C136 & a & 14.864\cite{adams1994wide} & 14.7417 & 14.7989 & 14.8817 & 14.6751\\
$\alpha$-Q & a & 4.9160 \cite{levien1980structure} & 5.0339 & 5.0280 & 4.9105 &  4.8872  \\
                & c & 5.4054 & 5.5216 & 5.5314 & 5.4022 & 5.3765 \\
$\beta$-Q & a & 4.9977 \cite{wright1981structure} & 5.1060 & 5.1171 & 5.05321 &  5.0529  \\
                & c & 5.4601 & 5.5869 & 5.5905 & 5.5207 & 5.52045 \\
$\alpha$-C & a & 4.9720 \cite{downs1994pressure} & 5.1026 & 5.1004 & 4.7418 &  4.8774  \\
                & c & 6.922 & 7.1364 & 7.1205 & 6.6199 &  6.8092 \\
$\beta$-C & a & 7.1590 \cite{barth1932cristobalite} & 7.4608 & 7.4625 & 7.5429 &  7.3724  \\
$\alpha$-T & a & 5.0100 \cite{cellai1995thermally} & 5.0616 & 5.1275 & 4.8502 &  4.9462  \\  
                & b & 8.600 & 8.9187 & 8.8062 & 8.3299 &  8.4948 \\
                & c & 8.2200 & 8.4166 & 8.4159 & 7.9607 &  8.1183 \\
$\beta$-T & a & 5.0350 \cite{villars1985pearson} & 5.2756 & 5.2731 & 5.2957 & 5.2255  \\
                & c & 8.220 & 8.6107 & 8.6069 & 8.6438 &  8.5292 \\ 
CO & a & 7.1356 \cite{levien1981high} & 7.2653 & 7.2354 & 7.2648 &  7.6759  \\
                & b &12.3692  & 12.5368 & 12.4253 & 11.9582 &  12.3520 \\
                & c & 7.1736 & 6.2766 & 6.2687 & 6.2943 &  6.6504 \\
SE & a & 4.0730 \cite{grocholski2013stability} & 4.1145 & 4.1155 & 4.2222 & 4.4444  \\
                & b & 5.0260 & 4.5263 & 4.5287 & 4.6461 & 4.8906 \\
                & c & 4.4770 & 5.0814 & 5.0830 & 5.2147 & 5.4892 \\
ST & a & 4.1797 \cite{ross1990high} & 4.2345 & 4.2285 & 4.2858 & 4.8368  \\
                & c & 2.6669 & 2.69170 & 2.6905 & 2.7269 &  3.0776 \\
KE & a & 7.464 \cite{shropshire1959crystal} & 7.5581 & 7.6446 & 7.4497 & 7.5001  \\
                & c & 8.62 & 9.1695 & 8.8285 & 8.6035 &  8.6617 \\
MO & a & 7.3317 \cite{miehe1992crystal} & 7.5199 & 7.4262 & 7.7782 &  7.4709  \\
                & b & 4.876 & 5.0728 & 4.9944 & 5.2312 &  5.0245 \\
                & c & 8.758 & 6.9780 &  6.8935 & 7.2203 &  6.9350 \\
CH & a & 13.6046 \cite{diaz1998synthesis} & 13.70564 & 13.7061 & 13.6346 & 13.5269  \\
                & c & 14.8290 & 14.9563 & 14.9569 & 14.8789 &  14.7614 \\
ZG & a & 16.4110 \cite{plevert2000gus} & 16.7106 & 16.9151 & 16.3306 &  16.6907  \\
                & b & 20.0440 & 20.4473 & 20.3998 & 19.6949 &  20.1292 \\
                & c & 5.0427 & 5.1559 &  5.2060 & 5.0261 &  5.1370 \\
ZSM & a &20.0511 \cite{artioli2000neutron}  &  20.4927 & 20.6177 &  20.2347 & 20.1282  \\
                & b &  19.8757 & 20.1528 & 20.2357 & 19.8598 &  19.7552 \\
                & c & 13.3682 & 13.5828 & 13.7043  & 13.4498 & 13.3789 \\
\shortstack{ RRMSE \\ (\%) }  & &  &  &0.17 &0.39 &0.37 \\
\hline
\end{tabular} 

\caption{\label{tab:L_param} Comparison of lattice constants predicted by MTP, ReaxFF, and COMB models against experimental data and DFT calculations.}
\end{table}

\begin{table} 

\begin{tabular}{lccccc}

\multicolumn{6}{c}{Si elastic constants (GPa)} \\ 
\hline
 & Exp \cite{brandes2013smithells} & DFT & MTP & COMB & ReaxFF\\
C$_{11}$&165&153.1&168.1&142.5&109.3\\
C$_{12}$&65&57.0&51.8&75.4&46.8\\
C$_{44}$&79.2&74.3&65.5&69.0&37.0\\
RRMSE (\%) &  &  & 5.7  & 8.0 & 21.5\\  
    &  &  & (5.8) & (7.0) & (18.8)\\  

 & \multicolumn{1}{l}{} & \multicolumn{1}{l}{} \\ 
\multicolumn{3}{c}{$\alpha-$quartz elastic constants (GPa)} \\ 
\hline
 & Exp \cite{mcskimin1965elastic} & DFT & MTP & BKS \cite{van1990force} \\
C$_{11}$  & 86.8  & 84.2 &  83.1&90.6\\
C$_{12}$ & 7   & 3.3  &  10.6&8.1\\
C$_{13}$  & 19.1  & 8.26 & 17.8&15.2\\
C$_{14}$  & -18  & 17.1 &  -21.8&-17.6\\
C$_{33}$ & 105.8   & 89.5  &  100.1&107.0\\
C$_{44}$  & 58.2  & 52.0 & 34.7&50.2\\
C$_{66}$  & 39.9  & 42.8 & 36.2&41.2\\
RRMSE (\%) &  &  & 6.1 & 2.4 \\ 
          &  &  & (12.3) & (10.8) \\  
\hline
\end{tabular} 

\caption{\label{tab:EConsts} Comparision of force fields prediction of elastic constant for silicon crystal and $\alpha$-quart. The RRMSE value calculated against DFT data are provided in parentheses. Note that the BKS potential was implemented by optimizing bulk parameters based on the predictions of experimental elastic constants of $\alpha$-quartz\cite{van1990force} } 

\end{table}

\begin{table} 
\centering
\begin{tabular}{lccccc}

\multicolumn{5}{c}{ } \\ 

Structure &Exp & DFT & MTP & COMB & ReaxFF \\
\hline
Diamond silicon & 98.3 \cite{brandes2013smithells} & 89.0 & 90.5 & 97.8 &  67.6  \\
$\beta$-tin silicon & & 105 & 110.0 & 180.3 & 98.2 \\
HD & & 87.3 & 91.2 & 97.8 &67\\
HCP silicon & &86.5  &  99.8 & 601.936 & - \\
BC8 silicon & & 83.4 &  94 &  95.8 & 291.1\\ 
ST12 silicon & & 75.0 &  99.5 & 56.9 &397\\ 
$\alpha$-quartz & 38 \cite{levien1980structure} & 32.8 & 39.9& 33.7 & - \\
$\alpha$-tridymite & & 26.1 & 22.6 & 20.8&-\\
Keatite & & 66.7& 70.9&  12.3& - \\
Coesite & 96.3 \cite{levien1981high} & 114 & 97.4 & - & -\\
Moganite & 32.2 \cite{leger2001high} & 37.2 & 28.9 & - & - \\
Seifertite & 290 \cite{pabst2013elastic} & 278 & 320 & 112.8 &  - \\
Zeolite (Chabazite) & 54 \cite{leardini2013elastic} & 62.0 & 56.3 & - &  -\\
Zeolite (GUS-1) & & 24.4 & 24.8 & 9.7 & - \\
Zeolite (ZSM) & & 32.9 & 47.1 & 11.5 & - \\
a-Si & - & 82.5\cite{durandurdu2001ab} & 70.9 & - & - \\
a-SiO2 & 36.4 \cite{guerette2012simple} & - & 37.9 & - & - \\
RRMSE &  &  & 3.3 & 51.4 & 44.7 \\
 &  &  & (3.5) & (24.1) & (33.6) \\
\hline
\end{tabular} 

\caption{\label{tab:Bulk_T} Exploring bulk modulus in silicon and silica polymorphs: Calculations using experimental lattice parameters input in lammps code. The units are expressed in GPa. The RRMSE value (calculated against DFT data) is provided in parentheses. The value of the bulk modulus for amorphous structures may vary depending on the cooling rate. In our study, we used cooling rates of 5 K/ps for silicon and 10 K/ps for silica.}
\end{table}

\begin{table} 

\begin{tabular}{lccc}

%\multicolumn{3}{c}{ Defect in silicon and silica (eV) } \\ 
%\hline
 & DFT & MTP & ReaxFF\\
 \hline
Si-V  & 3.7  & 2.3 &  3.0\\
Si-V$_{2}$  & 5.5  & 3.7 &  6.7\\
Si-I$_{2}$  & 2.9  & 3.0 &  1.9\\
Si-I$_{3}$  & 2.1  & 2.5 &  2.1\\
Si-I$_{4C1}$  & 2.2  & 2.5 &  2.0\\ 
Si-I$_{4C2}$  & 2.1  & 2.4 &  1.1\\
$\alpha$-AQ  & 6.1  & 5.8 &  6.6\\
$\beta$-BQ & 5.4   & 5.8  &  5.1\\
$\alpha$-AC  & 5.1  & 5.1 & 4.9\\
$\alpha$-AT & 5.0 & 4.8 & 1.3 \\
KE & 5.2   & 5.7  & 0.5 \\
CO  & 5.40  & 6.20 & 4.4 \\
CO & 5.9  & 6.78 & 4.3\\
MO & 5.2  & 5.8 & 5.5 \\
SE & 6.2  & 4.3 & 0.9\\
ST & 6.2 & 5.3 & 2.3\\
RRMSE (\%) &  & 4.5 & 12.0 \\ 
\hline
\end{tabular} 

\caption{\label{tab:Defects}Comparative analysis of defect formation energies in silicon crystals and silica polymorphs using DFT, MTP, and ReaxFF results. 
.Our comparison excludes the results from BKS and COMB as they are not parameterized for a single oxygen system.
In the training set, only the silicon vacancy defect is present, while other defects are used to test the portability of the potential. For Si-I$_{2}$, Si-I$_{3}$, Si-I$_{4C1}$, and Si-I$_{4C2}$, the formation energy is given in eV per atom.}
\end{table}

 The second-order elastic constants and bulk modulus are determined using finite difference as detailed in the methods section. Tab. \ref{tab:EConsts} provides the relative root mean square error (RRMSE) on elastic constant with respect to the DFT benchmark and experimental data. The MTP model demonstrates lower RRMSE when compared to those of ReaxFF and COMB potentials, it competes closely with the  Beest Kramer van Santen (BKS) \cite{van1990force} potential. Moreover, other semi-empirical models reported in Ref. \cite{bartok2018machine} demonstrate higher errors compared to the predictions made by the MTP model. The bulk modulus values for various silicon and silica polymorphs can also be found in Tab. \ref{tab:Bulk_T}. As evident, the MTP predictions closely align with the reference methods and experimental values, although it is worth noting that the training set did not encompass the deformation of certain polymorphs. Our potential accurately predicts the elastic constant of amorphous silica, even though amorphous configurations were not included in the training set. %\ref{fig:Dis}
In our testing of the MTP potential, we have also considered point defects like vacancies, divacancies, and self-interstitials. The RRMSE values for these defects are reported in Tab. \ref{tab:Defects}. Once again, using the MTP leads to smaller relative errors in comparison to ReaxFF potential. Since the majority of potentials were not specifically parameterized for the oxygen system alone, our comparison was limited solely to ReaxFF. In our study, we examined a specific case involving the I4 compact cluster \cite{arai1997self} within the Si crystal, which was not included in our training set. The atoms within the cluster exhibit a harmonious four-coordinated arrangement. Notably, the cluster boasts the presence of five-, six-, and seven-membered atomic rings. The bond lengths and bond angles of this cluster were calculated based on relaxed structures obtained from DFT, MTP, SW, ReaxFF and COMB calculations, as illustrated in Fig. \textcolor{blue}{2}. Notably, no dangling bonds were observed for all potentials, except for the COMB potential, which failed to reproduce the I4 structure. When analyzing the formation energy of the cluster, the MTP model exhibited a prediction within 14\% of the reference value, while the SW and ReaxFF potentials displayed errors reaching up to 27\% and 47\%, respectively. Among the models assessed, the MTP model exhibited superior agreement with DFT calculations for both bond lengths and bond angles.As this is a perfectly coordinated tetra-interstitial, we also tested 3 and 5-fold coordinated interstitials, namely, di-interstitial, tri-interstitial, and tetra-interstitial, as shown in Fig. \textcolor{blue}{2}. While these defects are not included in the training set, the MTP exhibits better agreement with the benchmark than ReaxFF, as detailed in Tab. \ref{tab:Defects}. The MTP also outperforms ReaxFF in describing the vacancy formation energy in silica polymorphs, as demonstrated in Tab. \ref{tab:Defects}. Again, no SiO$_{2}$ point defects were incorporated to our training set.
%\begin{comment}
\begin{figure}
    \centering
    \includegraphics[width=0.47\textwidth]{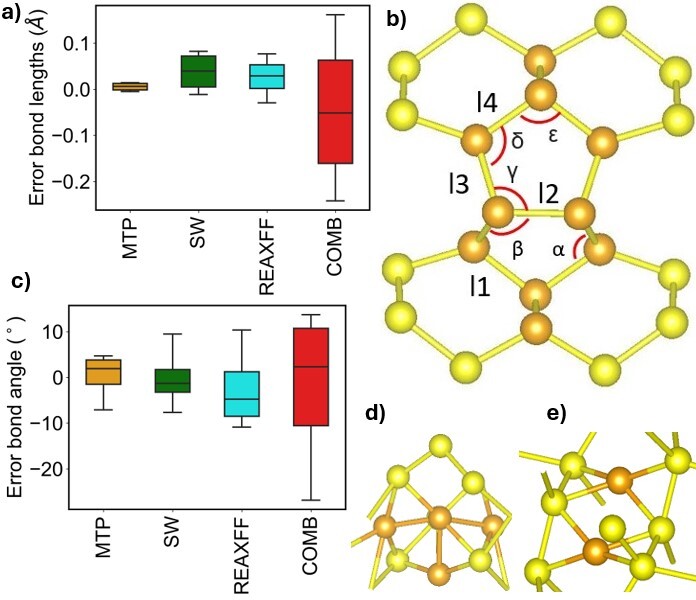}
    \caption{\label{fig:I4}\textbf{Point defects in silicon crystal.} In silicon crystal, a relaxed perfectly coordinated four-interstitial cluster is depicted. left \textbf{a} and \textbf{c}: the MTP leads to bond lengths and angles (as illustrated in the panel \textbf{b} ) in better agreement with DFT as compared to SW, ReaxFF and comb models. Additionally, two other interstitial clusters with coordination defects are presented, namely a four-interstitial cluster \textbf{d} and a di-interstitial \textbf{e}. All the interstitial atoms are colored orange . In \textbf{d} and \textbf{e}, the interstitial bonds are also colored orange. Their formation energies are indicated in Tab. \ref{tab:Defects}, where configurations \textbf{b}, \textbf{d}, and \textbf{e} correspond to Si-I${4C2}$, Si-I${4C1}$, and Si-I$_{2}$, respectively. The notations {\sffamily I$_{1}$, \sffamily I$_{2}$, \sffamily I$_{3}$}, and {\sffamily I$_{4}$}, as well as $\alpha$, $\beta$, $\gamma$, $\delta$, and $\epsilon$, represent bond lengths and bond angles within the cluster, respectively. These interstitial configurations were not included in the training set.}
    %\label{fig:foobar}
\end{figure}
%\end{comment}

The static migration barrier energy of the vacancy was determined using the nudged elastic band (NEB) \cite{henkelman2000climbing}. The migration barrier profiles, including the DFT-based profiles as well as the MTP- and SW-based profiles, are depicted in Fig. \textcolor{blue}{3}b.The MTP migration barrier profile shows excellent agreement with the DFT reference profile. In contrast, the SW potential does not capture the reference profiles with similar accuracy. Examining the barrier for vacancy migration reveals a relative error in barrier height of 15.0\% for MTP, while for SW, it is 73.1\%.
%\begin{comment}
\begin{figure}
\vspace*{-1cm}
\centering
    \includegraphics[width=7.25cm]{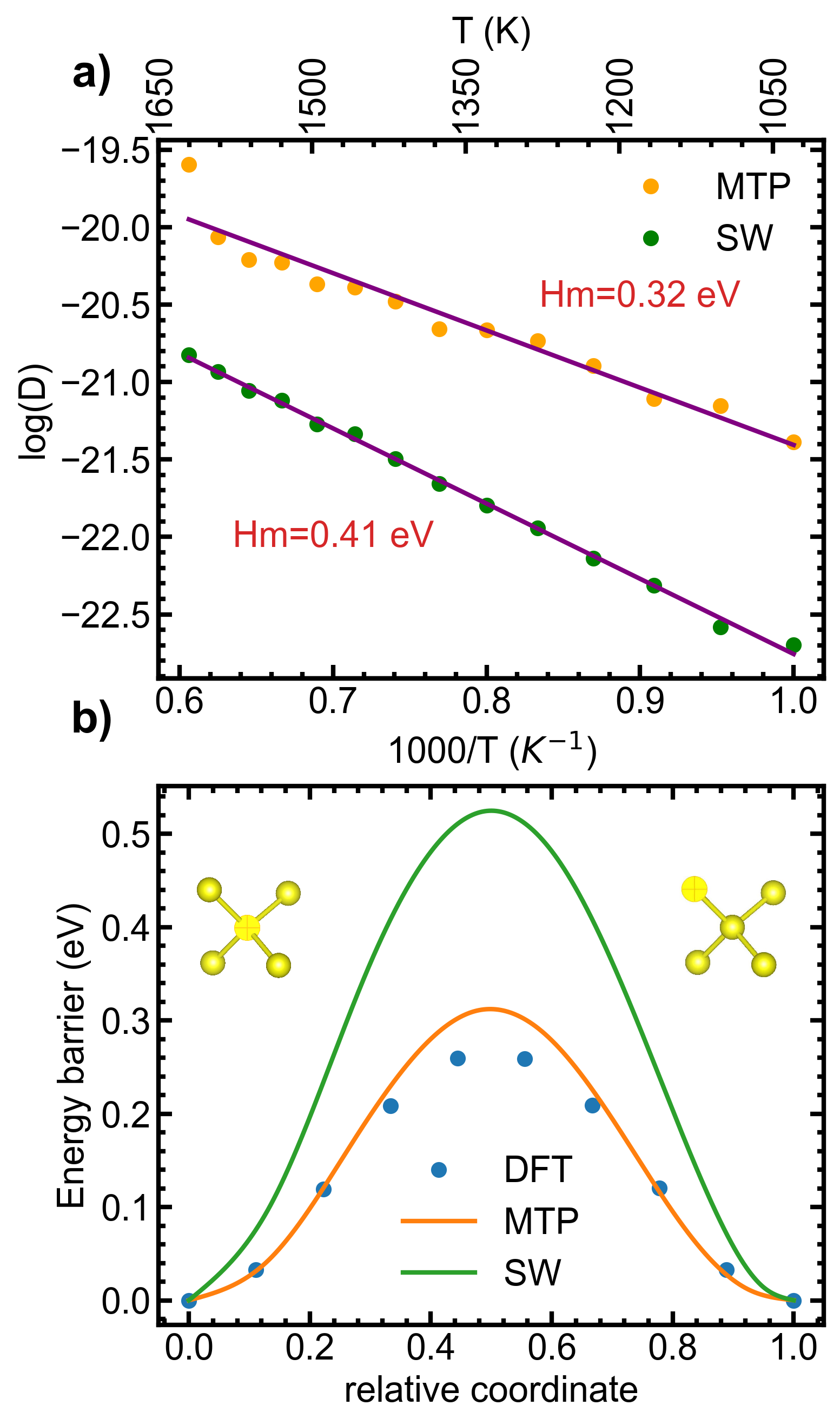}
    \caption{\label{fig:PointDefects}\textbf{Diffusion of point defects in silicon crystal}.
\textbf{a} Temperature dependence of vacancy diffusion coefficients simulated using the MTP and SW. \textbf{b} NEB-based mono-vacancy jump barrier. DFT, MTP and SW are compared. AIMD configurations containing vacancies, along with NEB configurations, were excluded from the training set. Insets: Illustration of vacancy position before and after the jump within the silicon crystal.}
\end{figure}
%\end{comment}
These results demonstrate that the MTP model can better mimic the reference method when studying point defects within larger systems, as demonstrated in references \cite{ohbitsu2022atomic, dragoni2018achieving, cheng2020vacancy}. 
The activation barriers for mono-vacancy hopping were further investigated using molecular dynamic simulations, considering temperatures ranging from 1000 K to 1650 K. The simulation details are provided in the method section for reference. The mean square displacements are also provided in the supplementary Fig. S4. As reported in Fig. \textcolor{blue}{3}a, the activation energy for mono-vacancy hopping is 0.31 eV for MTP and 0.41 eV for SW. These values are close to the static activation barriers computed at 0 K; the MTP model behaves in a physically plausible fashion. As observed in Fig. \textcolor{blue}{3}a-b, the barrier obtained from MD simulation is 0.32 eV, whereas the static migration barrier stands at 0.31 eV. It is noteworthy that NEB configurations, including AIMD configurations containing vacancies, were not included in the training set. We investigated extended defects such as generalized stacking faults and dislocations in the Si crystal. Generalized stacking faults are planar defects closely linked to slip. In turn, the behavior of dislocations and their core properties are particularly important for understanding plasticity. The methods section of our study provides a detailed description of our model for generalized stacking faults and dislocations, as well as the calculations involved. In Fig. \textcolor{blue}{4}a-b, the excess energy per unit area, also known as a $\gamma$-line, is presented. The $\gamma$-line was calculated using the benchmark method, the MTP model, the SW \cite{stillinger1985computer}  and Tersoff (TS) \cite{{tersoff1986new},{tersoff1988empirical}} semi-classical potentials. As shown in the Fig. \textcolor{blue}{4}, the MTP is in good agreement with the benchmark results. In contrast, SW and TS demonstrate lesser agreement with the benchmark. In Fig. \textcolor{blue}{5}, the dislocation core structures are presented. The core structures predicted by the MTP model exhibit a nearly perfect agreement with those predicted by DFT, as reported in Ref. \cite{guenole2010determination}. An important and distinctive feature of our potential is the direct relaxation of the C$_{1}$ core structure to the C$_{2}$ structure, which is commonly referred to as the double-period reconstruction of the C$_{1}$ core. In past studies, the C$_{2}$ core structure was often manually reconstructed from the relaxed C$_{1}$ core \cite{huang2021locking,guenole2010determination}. However, our potential eliminates the need for manual reconstruction by obtaining the C$_{2}$ structure directly. To obtain the relaxed structure and energy of the C$_{1}$ configuration, a snapshot is selected from the relaxation steps that ultimately lead to the C$_{2}$ structure. Our investigation also revealed that the C$_{2}$ core is the most stable configuration, a result consistent with previous reports \cite{huang2021locking,guenole2010determination}. In addition to Fig. \textcolor{blue}{5}, the core structures are also depicted in Supplementary Fig. S10.
\subsection*{Coexistence simulation}
We determined the solid-liquid coexistence temperature of silicon using the solid-liquid interface method described in Ref. \cite{morris1994melting}. Our MTP potential predicts the silicon melting point to be 1485 ± 5 K, which is about 0.5\% lower than the benchmark DFT-GGA value of 1492 K \cite{alfe2003exchange}. Note that both the MTP potential and the benchmark value are approximately 12\% lower than the experimental melting point of 1687 K, as reported in Ref. \cite{lide2004crc}. Notably, our database initially lacked solid-liquid interface data, so we integrated a few AIMD configurations gathered around the experimental melting point into our unified training set. Additionally, it is important to acknowledge the influence of the exchange-correlation (XC) functional on melting behavior, as discussed in Ref. \cite{alfe2003exchange, dorner2018melting}. The capability of MTP to accurately replicate the melting point of the GGA exchange-correlation (XC) functional showcases the high-quality simulation of liquid structure by MTP, as elaborated in the following section. For comprehensive details on our coexistence simulation approach, please refer to the accompanying supplementary Fig. S9.
%\begin{comment}
\begin{figure}
\centering
    \includegraphics[width=8.5cm]{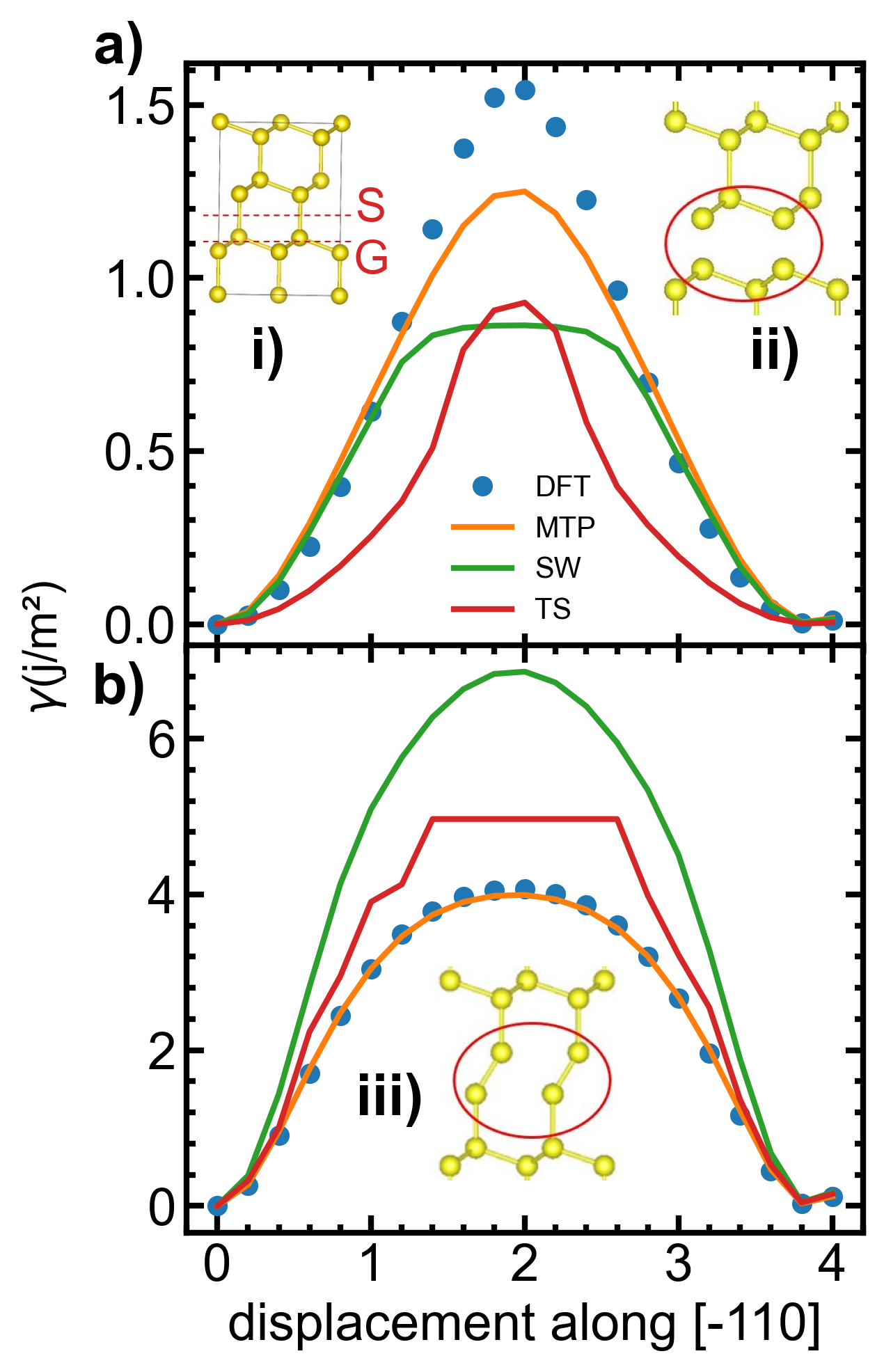}
    \caption{\label{fig:GSFE}\textbf{Planar defect in silicon crystal.} $\gamma$-lines on the (111) plane as predicted by DFT, MTP, TS and SW. The shuffle (S) and glide (G) cuts are illustrated in the inset. The MTP provides a good description of the $\gamma$-line associated to the shuffle cut (top) and a near-perfect description of the $\gamma$-line associated to the glide cut. Insets i), ii), and iii) represent the bulk structure, the relaxed shuffle (S) structure, and the relaxed glide (G) structure, respectively.}
\end{figure}

\begin{figure}
\centering
    \includegraphics[width=8.5cm]{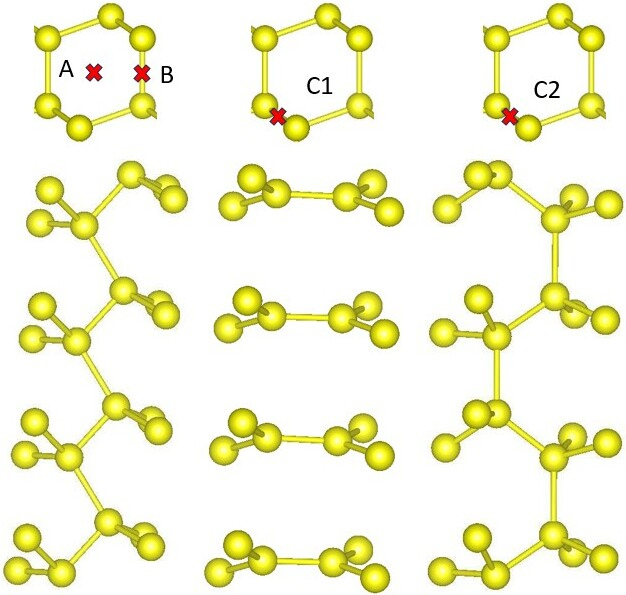}
    \caption{\label{fig:Dislocation}\textbf{Line defect in silicon crystal.} Positions and relaxed structures of [110] screw dislocation cores in Si obtained with the MTP potential. The system size was set to 14400 atoms and oriented such that the x, y and z  directions coincide to [11-2], [111], [110]  respectively. The dislocation core structures are in good agreement with DFT core structures reported in the literature\cite{guenole2010determination}. The core types are represented by A, B, C$_{1}$, and C$_{2}$ respectively. The red mark indicates the position of the dislocation core. Dislocation configurations were excluded from the training dataset.}
\end{figure}
%\end{comment}

\subsection*{Silicon slab energy }
We compare the surface energy predicted by MTP against experimental data and DFT references, as slab data were not included in the training set. Experimental slab energy values for Si (100), Si(110), and Si(111) are reported to be 2.1, 1.5, and 1.2 j/m$^{2}$ \cite{jaccodine1963surface}, respectively, while DFT values are 2.1, 1.8, and 1.6. The MTP predicts these surface energies to be 2.0, 1.3, and 1.2, respectively. While a large discrepancy between MTP and DFT is observed, especially for Si(110) with relative errors of up to 25$\%$, the MTP-predicted surface energy closely matches the experimental slab energy.
\subsection*{Disordered structures}
We also considered an extensive set of disordered Si and SiO$_2$ structures. We compared structures generated using \textit{ab initio} MD, MTP-MD, and semi-empirical-MD. All three cases were subjected to identical MD simulation conditions, except for amorphous silica, where the MTP simulation time was shorter compared to the other potentials. For comparison, we have chosen the Vashista (VA) \cite{vashishta1990interaction}, Munetoh (TS) \cite{munetoh2007interatomic}, BKS \cite{van1990force}, and Sundararaman (SHK1 and SHK2) \cite{sundararaman2018new} models. 
The details regarding the \textit{ab initio} and classical MD simulations are comprehensively provided in the methods section. 
To analyze the disordered structures, we utilized both the pair correlation functions and the bond angle distribution functions.
\subsection*{Liquid and amorphous silicon}
As observed from the radial distribution function shown in Fig. \textcolor{blue}{6}a, the MTP describes the structure of liquid Si with high precision. In contrast, the others potential leads to a shifted position of the first-neighbor peak compared to the results obtained from the reference (DFT). Additionally, there is an overestimation of the peak height, primarily observed with the EDIP and Tersoff potentials.
The angular distribution function Fig. \textcolor{blue}{6}b also demonstrates excellent agreement between the MTP model and experimental data.
When considering amorphous Si, the MTP model accurately describes the structural features in agreement with experimental data. Conversely, semi-classical potentials like SW and Tersoff fail to replicate the experimental radial distribution profile.
As shown in Fig. \textcolor{blue}{6}c, experimental g(r) and MTP-MD lead to nearly identical first-neighbor peaks (2.36 Å) and second neighbor peaks (3.89 Å). 
Additionally, the MTP model exhibits better agreement with the experimental bond angle distribution of 108.6° \cite{fortner1989radial}. Although the bond angle distributions of semi-empirical models like SW and EDIP are closer to the experimental values, they exhibit angle distributions below 90°, as shown in Fig. \textcolor{blue}{6}d.
Note that the \textit{ab initio} cooling and equilibration trajectory was not included in the training set of the MTP model, which suggests the MTP is fairly general, which can be attributed to the fact that the training dataset encompasses a variety of configurations within the disordered Si systems.
Overall, the MTP model demonstrates a level of accuracy comparable to that of DFT and experiment when describing the structural features and bonding characteristics of disordered Si. 
%\begin{comment}
\begin{figure}
\vspace*{-1cm}
\centering
    \includegraphics[width=8cm]{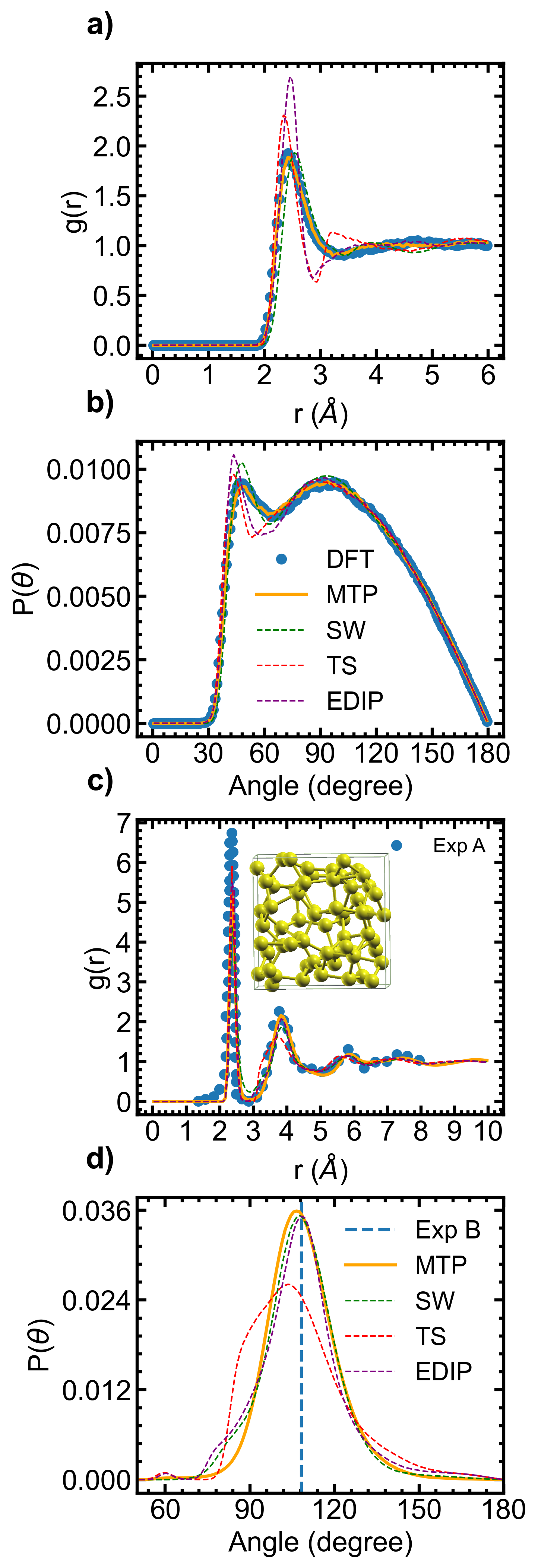}
    \caption{\label{fig:SiRDF}\textbf{Liquid and amorphous silicon.} Disordered structures of Si simulated using DFT, MTP, semi-empirical models \textbf{a} Radial and \textbf{b} angular distribution functions of liquid Si (3370K, 64 atoms); \textbf{c} Radial and \textbf{d} angular distribution functions of amorphous Si (300 K, 1000 atoms). The amorphous distribution functions are compared against experimental data from Exp A \cite{laaziri1999high, meidanshahi2019electronic} and Exp B \cite{fortner1989radial}. }
\end{figure}
\begin{figure*}
    \centering
    \includegraphics[width=18cm]{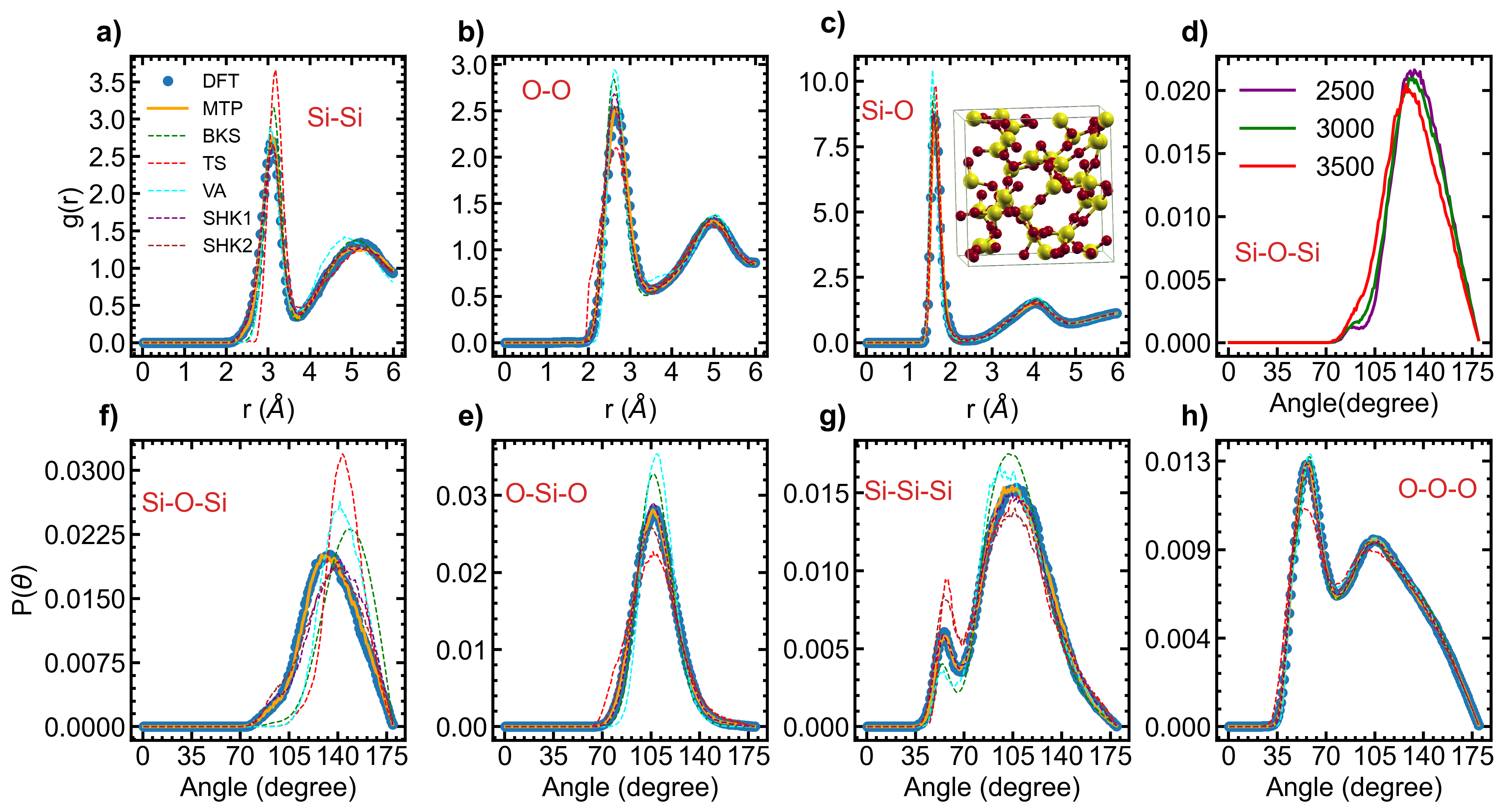}
    \caption{\label{fig:SilicaRDF}\textbf{Liquid silica.} Si-Si, O-O and Si-O correlation functions g(r) and partial bond angle distribution function
for liquid SiO$_2$ at 3600 K simulated using DFT, MTP, BKS, TS, VA, SHK1 and SHK2 with a 96-atom simulation box. Top-right: Distribution of the Si-O-Si angle against temperature using the MTP potential.}
\bigskip
\end{figure*}
%\end{comment}
\subsection*{Liquid and amorphous silica}
Fig. \textcolor{blue}{7} illustrates results pertaining to liquid SiO$_2$. It includes pair distribution functions (PDF)and angular distribution functions (ADF).
The MTP is in better agreement with DFT as compared to the others potential.
We notice both quantitative and qualitative differences between the semi-empirical potentials and DFT, except for the Si-O pair correlation function. In this case, the semi-empirical potentials only overestimates the height of the first peak, which is located around 1.62 Å \cite{mozzi1969structure, grimley1990neutron,mei2007structure}. This value is consistent with the experimental Si-O bond length observed in liquid SiO$_2$, indicating a strong chemical interaction between the Si-O pairs.
Both the O-O and Si-Si pair correlation functions exhibit a shift in the first peak as given by the others models, while there is a strong quantitative and qualitative match between the MTP-based and the DFT-based structures.
Furthermore, the BKS-based Si-O-Si ADF, along with those of other semi-empirical models, does not match the DFT-based ADF. Conversely, the MTP-based ADF demonstrates a good match.
At 3600 K, using our 96-atom simulation box model, the average Si-O-Si angle between two SiO$_4$ tetrahedra is determined as follows: DFT - 134.5°, MTP - 131.4°, BKS - 146.0°, TS - 143.6°, VA - 141.8°, SHK1 - 142.6°, and SHK2 - 140.4°. Our DFT value aligns closely with literature, approximately 136.0 \cite{carre2016developing} and 135.0 \cite{liu2019balance} respectively. It's evident that the MTP closely resembles the reference method, whereas other models align with experimental values for the Si-O-Si angle in amorphous silica, ranging between 140° and 152° \cite{farnan1992quantification, da1975refinement, coombs1985nature} . This likely stems from the semi-empirical models being meticulously fitted with consideration of experimental properties. Most of them were optimized based on mixed ab initio-experimental data. We then varied the temperature of liquid silica from 2500 K to 3500 K using a 648-atom box and recorded the Si-O-Si angle, as depicted in Fig. \textcolor{blue}{7}d. We have found that the Si-O-Si angle in liquid silica changes with temperature, consistent with the findings reported in Ref. \cite{kobayashi2021self}. As the temperature decreases, the angle between tetrahedral interconnections increases, contributing to network relaxation. It is likely that the relaxation of the network at room temperature upon cooling is primarily attributed to variations in bond lengths and angles, given that the network structure of silica liquid does not qualitatively change between 3500 K and 300 K.

Furthermore\textit{Ab initio} MD, MTP-MD, BKS-MD and other models lead to a within-tetrahedra O-Si-O bond angle distribution centered around 109°, which is nearly equal to the experimental bond angle\cite{mozzi1969structure, grimley1990neutron,mei2007structure}. However, the BKS and VA potentials overestimate the average probability at 109°, while the TS potential underestimates this probability. The SHK1, SHK2, and MTP potentials match the benchmark. 
The DFT, MTP, and all other models except TS lead to very similar O-O-O ADFs. However, when it comes to the Si-Si-Si angle, qualitative and quantitative discrepancies are observed between BKS-generated structures and DFT-generated structures, while MTP-generated structures match the benchmark. There is also a notable discrepancy between the structure predicted by other models and that of the benchmark.
%\begin{comment}
\begin{figure*} 
\centering
    %includegraphics[width=8.5cm]{AM_SIO_RG.png}
    \includegraphics[width=18cm]{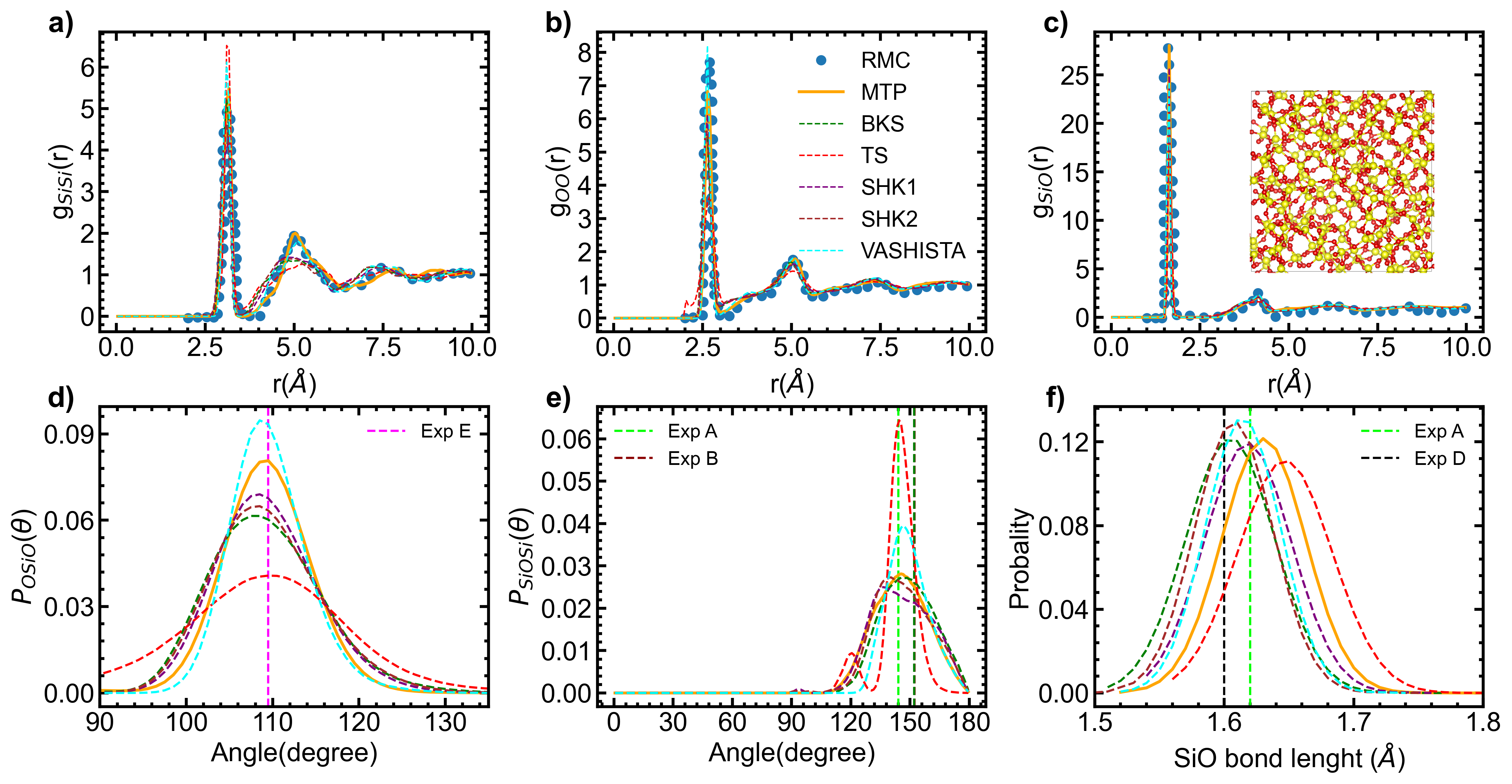}
    \caption{\label{fig:GlassySilica}\textbf{Amorphous silica.} Partial radial distribution functions in vitreous silica were obtained using various potential including MTP, BKS, TS, VA, SHK1, and SHK2. The vitreous systems were equilibrated at 300 K, consisting of 648 atoms. These partial radial distribution functions are compared with experimental data obtained using Reverse Monte Carlo (RMC) \cite{tucker2005refinement}. Additionally, the angular distribution function and bond length distribution function were analyzed and compared with experimental data from multiple sources: Exp A \cite{mozzi1969structure}, Exp B \cite{da1975refinement}, Exp C \cite{vukcevich1972new}, and Exp D \cite{khouchaf2020study} }. 
\end{figure*}
%\end{comment}

The partial pair distribution functions of vitreous SiO$_2$ are illustrated in Fig. \textcolor{blue}{8}a-c. The comparison is made against the pair distribution function computed from experimental data using the reverse Monte Carlo (RMC) method \cite{tucker2005refinement}. The MTP potential exhibits qualitative agreement with RMC data, as our pair distribution function profiles match those obtained from RMC.For example, only the MTP accurately reproduces the RMC profiles for Si-Si interactions, as the second peak around 5 Å is not reproduced by the other models, including the BKS model. Additionally, while the other models fail to reproduce the height of the Si-O RMC pair distribution function, the MTP potential shows a good match. To further analyze the structure, we also computed the most important angular distribution function, as well as the Si-O bond length distribution function. For O-Si-O (refer to Fig. \textcolor{blue}{8}c, all the profiles are similar but vary in height, centered around the experimental value of 109.0°.  When it comes to Si-O-Si angle ( Fig. \textcolor{blue}{8}d ), both the MTP and BKS show similar profiles centered between the experimental values of 144° and 152°. The average values for BKS Si-O-Si angle is 150 and the one of the MTP is 145.5. As the average value for the same angle predicted by MTP in liquid silica at 3600 is around 131.4, this confirms that the Si-O-Si angle varies in liquid silica. Considering all the potential, the the Si-O bond length distribution are centered between 1.60 and 1.66 \AA. The average bond length distribution for MTP potential is 1.63 \AA which is close to experimental values of 1.62 \AA. Note that these structural properties may vary slightly depending on the employed cooling rate.
Despite the high cooling rate, no major coordination defects were observed. This indicates that the configuration has been well equilibrated at 3000 K, resulting in the establishment of a strong network. Note that the \textit{ab initio} MD trajectory, which encompasses both the cooling and equilibration stages of the amorphous structure preparation, was not included in the MTP training set, which is an indicator of the MTP's generalization capability. 
Overall, the MTP potential demonstrates a remarkable improvement in accurately describing the structure of disordered SiO$_2$ compared to well-established potentials such as the BKS potential and others. The MTP model captures the essential structural features of the disordered systems with greater precision, resulting in a better agreement with experimental observations\cite{mozzi1969structure,grimley1990neutron, tucker2005refinement,mei2007structure}.
%\begin{comment}
\begin{figure}
\centering
    \includegraphics[width=8.5cm]{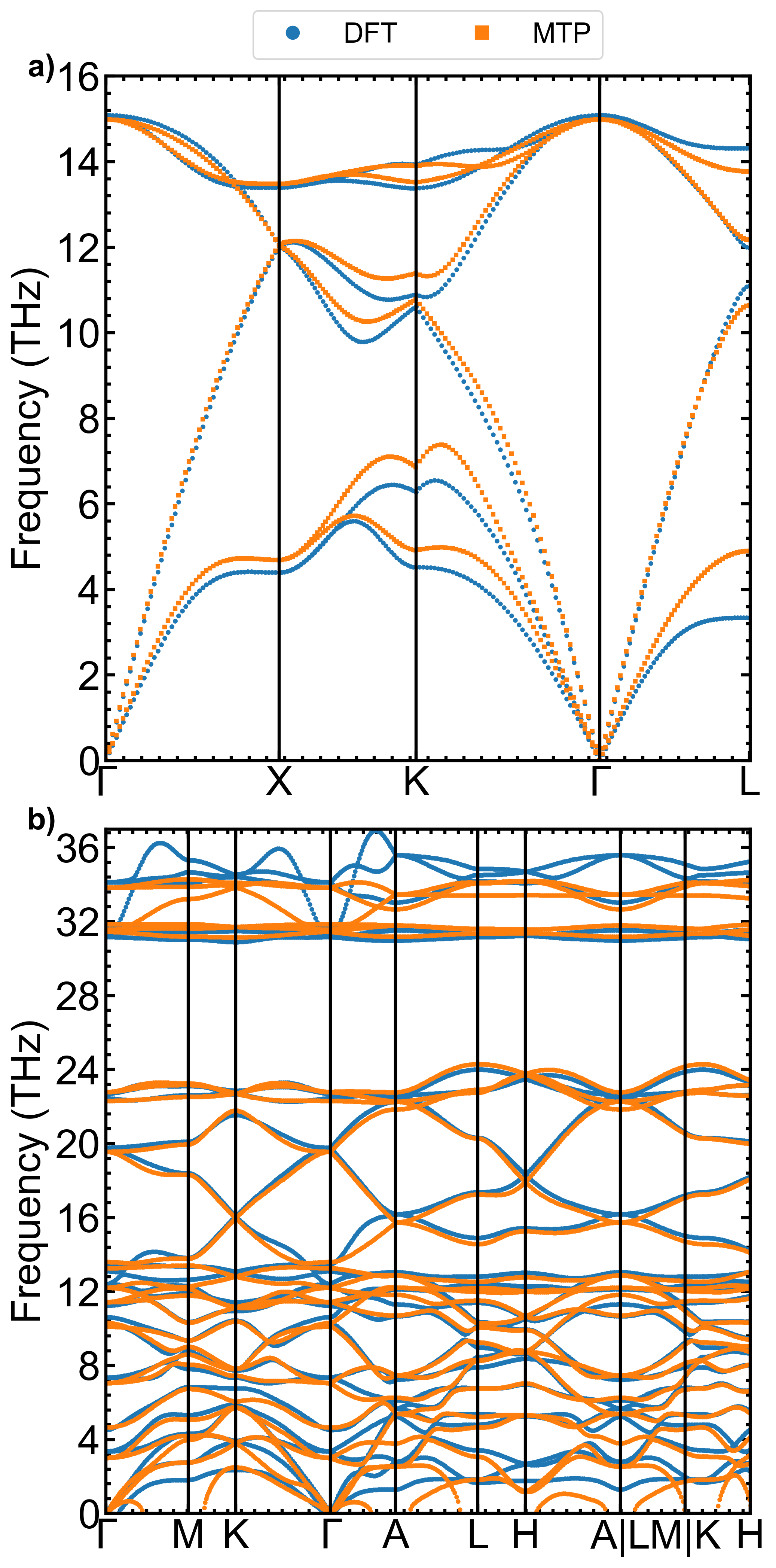}
    \caption{\label{fig:Phonons}\textbf{Phonon dispersions.} Phonon dispersions of \textbf{a} \textit{c}-Si and \textbf{b} $\alpha$-quartz computed using DFT and MTP.}
\end{figure}
%\end{comment}
\subsection*{Phonon dispersion}
We calculated the phonon dispersion of c-Si and $\alpha$-quartz, as illustrated in Fig. \textcolor{blue}{9}a-b respectively. The MTP model exhibits very good agreement with the reference method (DFT) except the higher frequencies for  $\alpha$-quartz. Once again, the corresponding frozen phonon configurations were not explicitly included in the training set. 

\subsection*{Si - SiO$_{2}$ Interface} 
The Si-SiO${_2}$ interface, a cornerstone in semiconductor physics and material science, plays a fundamental role in device fabrication and significantly impacts device performance. Exploring hetero-structures involving Si - SiO$_{2}$ interfaces opens avenues for novel functionalities and applications in microelectronics and beyond. Given its importance, capturing the structure and dynamics of the Si - SiO$_{2}$ interface is paramount for a potential model of the Si-O system. To achieve this, we employed various models, encompassing crystalline Si slabs with different orientations, such as Si(100), Si(110), and Si(111). Our approach involved utilizing both $\alpha$-quartz, $\beta$-cristobalite, and other polymorphs in constructing the Si - SiO$_{2}$ interface. For a detailed explanation of our construction scheme, please refer to the Method section. 
To assess the suitability of the potential for modeling the Si/SiO$_{2}$ interface, we examined interfaces using both O-terminated SiO$_{2}$ slabs and Si-terminated slabs.
For each Si/SiO$_{2}$ crystalline interface configuration, we perform force and energy minimization, allowing the positions of atoms and the simulation box size to change simultaneously. Following geometry optimization, molecular dynamics simulations are conducted for 50 ps at 300 K in the NPT ensemble.  First, our potential stabilized the majority of the interfaces in both static and dynamic runs, with the extent of stabilization depending on the orientation of the Si slab and the termination of the quartz slab. For Si(110) in contact with either a Si- or O-terminated quartz slab, our potential successfully stabilizes and describes the dynamics of the resulting interfaces, whether symmetric or non-symmetric. However, our potential only successfully describes symmetric interfaces for Si(100) and Si(111), which are built from quartz slabs terminated by Si. These combinations of termination and orientation were not considered in the training set, showcasing the remarkable generalization ability of our interaction potential. It's worth noting that defects, such as silicon dangling bonds or over-coordinated oxygen atoms, are observed at the interface following both minimization and dynamic runs, as depicted in Fig. \textcolor{blue}{10}. As noted in Ref. \cite{himpsel1988microscopic}, the presence of the dangling bonds is a natural occurrence and constitutes a typical aspect of interface defects. Such anomalies are commonplace and anticipated in these interfaces, owing to the inherent lattice mismatch between the involved materials. Usually, interfacial defects are passivated or special construction schemes are adopted to eliminate them. However, our study does not aim to create defect-free interfaces. Instead, our goal is to evaluate the potential's capability to manage complex heterostructures with varying bonding types not encountered during the training process. 
%\textcolor{red}{
To further validate our potential for silicon and silica interfaces, we compare the interfacial energies of small models computed using MTP and DFT. Detailed descriptions of these small models are provided in the Methods section. Our results (shown in Fig. \textcolor{blue}{11}) exhibit good correlation between DFT and MTP models. Importantly, configurations generated by the MTP potential through relaxation and MD simulations converged easily, typically requiring fewer than 50 iterations of single-point energy calculations of DFT. This highlights the reliability of interfacial configurations generated by the MTP potential. These findings demonstrate the capability of MTP potentials to effectively investigate heterosystems containing Si-SiO$_2$ interfaces. The relaxed small models of the Si-SiO$_{2}$ interfaces are presented in the supporting information (Fig. S11 and Fig. S12).
%\textcolor{red}{
% \begin{comment}
\begin{figure}
    \centering
    \includegraphics[width=0.48\textwidth]{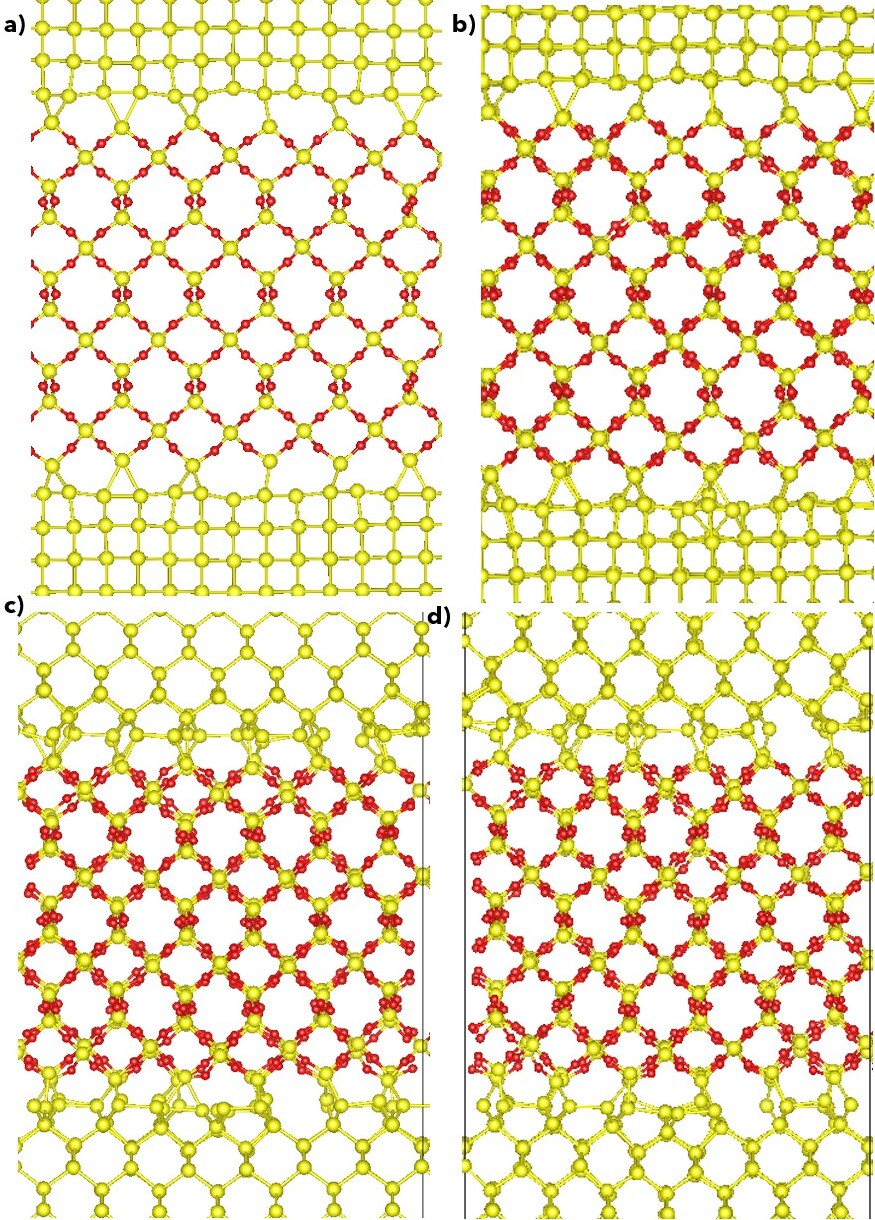}
    \caption{\label{fig:INT}\textbf{Interface structure of Si - SiO$_{2}$.} Geometry optimization and MD equilibration of 50 ps simulation:
Si(010)/$\alpha$-Quartz (001) (top) and Si(110)/$\alpha$-Quartz (001) (bottom)
\textbf{a} $\&$ \textbf{c} Relaxed at 0 K, \textbf{b} $\&$ \textbf{d} Annealed at 300 K}.
\end{figure} 
%\end{comment}

%\begin{comment}
\begin{figure}
    \centering
    \includegraphics[width=0.48\textwidth]{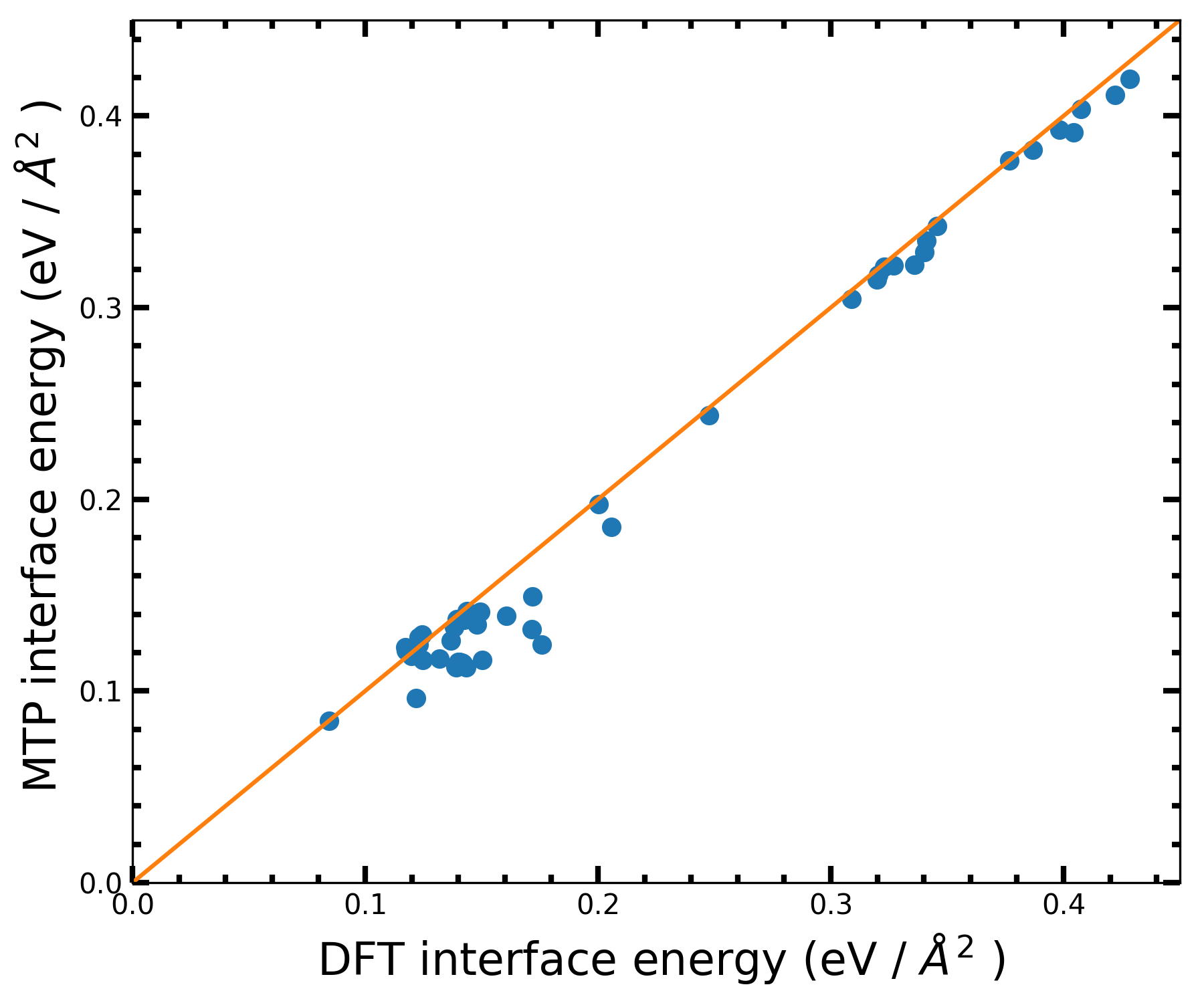}
    \caption{\label{fig:INTERF_E}\textbf{Interface energy of Si-SiO$_{2}$.} Comparing interface energies of silicon and various silica polymorphs, including amorphous silica, using MTP potential and DFT. The orange line indicates the diagonal y = x, corresponding to a perfect correlation.}
\end{figure} 
%\end{comment}

\section*{Discussion} 

In this work, we have successfully parametrized a machine learning (ML) potential that can implicitly capture and describe different charge states. Additionally, this ML potential has the remarkable ability to describe disjoint zones and hetero-zones of the configurational space of SiO$_2$ and its constituent elements, Si and oxygen.
The potential description of various phenomena—including point defects, diffusion in Si crystals, extended defects, the liquid phase of Si and SiO$_2$, and the amorphous phase of Si and SiO$_2$—either rivals or out-performs existing potentials.
The potential exhibits very good agreement with experimental data in challenging configurational zones, such as the amorphous state, even though these configurations were not included in the training data.
In many scenarios, such as disordered phases (liquid, amorphous), the potential achieved a near-perfect match with the reference method in terms of accuracy, using exactly the same simulation time(very short) and conditions. Even when utilizing longer simulation times and large system with a semi-empirical model, it does not reaches a level of accuracy similar to the reference method. Furthermore, the potential displays an intriguing capability regarding dislocation behavior. It autonomously transitioned the C$_{1}$ structure to the C$_{2}$ structure without the need for manual reconstruction.
While there is a growing consensus that machine learning (ML) potentials can effectively serve as surrogates for density functional theory (DFT) in terms of accuracy and speed, generalization including charge state modeling remains a challenging task. This study provides evidence that reaction coordinates are sufficient to implicitly capture charge transfer or charge states involved in chemical reactions.  
Aside from potentials explicitly considering charge-transfer, separate machine learning (ML) potentials are developed for each individual chemical element or compound which database is constructible by DFT.
Our approach suggests this is not necessary; compounds and their elemental constituents can be trained jointly. Indeed, joint parametrization, where parameters are derived simultaneously for both silicon, silica and oxygen, offers several advantages. By employing a single potential to describe the interactions between atoms in both silicon and silica, the computational model becomes more streamlined and easier to manage. The unified potential can save time and resources by avoiding the need to recalibrate parameters of the model for each of the materials involved. The unification idea is also important for some areas of application, such as interfacial modeling for electronic devices, energy storage and conversion, and surface coatings and tribology. Here, the reference data for machine learning potential must include both the individual materials in contact as well as the boundary region. In addition, as chemical reactions can occur in MD simulation, good modeling of a multi-component material or complex systems under certain conditions requires joint parametrization of the considered system and its constituents. For instance, oxygen aggregation in high-temperature molecular dynamics (MD) simulations was observed by researchers, as noted in Ref. \cite{erhard2022machine} (supporting information). This observation led to the inclusion of oxygen molecules in the training set by the researchers. 
Our preliminary study pertained to a semiconductor and its oxide. However, whether this approach can be generalized to other elements and mixtures—including multi-component alloys and compounds remains to be seen. To ensure a well-implemented unified interaction potential, several other aspects need to be explored in the future. Given that the compound and its constituents are not located within the same zone of the configurational space, achieving an accurate, efficient, low-cost, and general unified potential for both the material and its oxides with limited data may require adjustments to the underlying mathematical model, the fitting procedure, and the database sampling methods (including active learning). These adjustments could help attain the same level of accuracy at a more affordable computational cost, resembling a feature of potentials parameterized for a single compound.
Likewise, while this study achieved a joint description of Si and SiO$_2$ using the MTP framework, it is likely that other currently developed ML potential frameworks would have led to a similar result.
In conjunction with these questions, our aim in the future is to extend this work by incorporating the element of hydrogen to model silica gels.

\fontsize{11}{12}\selectfont

\section*{Methods}

\subsection*{\textit{Ab initio} calculations}
\noindent
The database was constructed using DFT, as implemented in the Quantum ESPRESSO \cite{giannozzi2009quantum} package.
The exchange-correlation potential was treated using the generalized gradient approximation of Perdew-Burke-Ernzerhof (GGA-PBE) \cite{perdew1996generalized}.
Projector augmented waves (PAW) \cite{blochl1994projector} were employed.
Kinetic energy cutoffs of 884 eV for Si and 1224 eV for both SiO$_2$ and oxygen were chosen.
In all calculations, the Brillouin zone was sampled using the Monkhorst-Pack grid \cite{monkhorst1976special} scheme.
Different k-points were used for each polymorph, including an 8x8x8 for the ordinary phases of Si and an 11x11x11 for SiO$_2$. The gamma point was used for oxygen molecules.
\subsection*{Machine learning model: The moment tensor potential}
\noindent
In this work, the moment tensor potential (MTP) \cite{shapeev2016moment} was chosen as the machine learning model.
The moment tensor potential (MTP) is a multi-component potential. In a previous comparative study \cite{zuo2020performance}, it demonstrated a favorable trade-off between accuracy and computational speed across a range of modeling problems. The model derives its name from its use of a tensorial representation of atomic coordinates and utilizes linear regression to determine the local atomic energy. These local atomic energies are subsequently summed to obtain the total energy of the system under consideration. The MTP model considers the total energy of a specific atomic configuration as a sum of individual atomic energy contributions.
\begin{equation} \tag{1}
E_{Total}=\sum_{i=1}^{n}E_{i}=\sum_{i=1}^{n}V_{local}(\zeta_{i}) 
\end{equation}
The argument $\zeta_{i}$ is a tuple $\zeta_{i} = (r_{ij}, \tau_i, \tau_j)$ containing the relative coordinate $r_{ij}$ and atomic types $\tau_i, \tau_j$. Here, V$_{local}$ is approximately computed within the sphere or circle of radius ( Rc) of 5.7 \AA, beyond which the central atom no longer feels any interaction.
Practically, in the MTP framework, the expansion of the atomic energy $V_{local}$ into basis functions $B_\beta$ serves as the foundation for linear regression. within the sphere or circle of radius
\begin{equation} \tag{2}
V_{local} =\sum_\beta c_{\beta}B_\beta
\end{equation} 
Since the potential energy function $V_{local}$ is smooth, the force acting on an atom $k$ at position $r_k$ in a given configuration $x_q$ can be calculated by taking the gradient of the total energy.
\begin{equation}\tag{3}
F_k(x_q)=-\nabla E_{Total}(x_q)=-\sum_{i}\frac{\partial V(\zeta_{i})}{\partial r_k(x_q)}
\end{equation}
The virial stress within an atomic configuration $x_q$ of volume $\Omega$ can be expressed as follows.
\begin{equation}\tag{4}
\sigma_{ij}(x_q)=\frac{1}{2\Omega}\sum_{k \in \Omega}\sum_{l \in \Omega}(x_i^{(l)}-x_i^{(k)})F_{j}^{(kl)}
\end{equation}
The functions $B_\beta$ in equation 2 are obtained through the contraction of the descriptors. In the MTP model, the descriptors are formed by tensors of atomic coordinates weighted by radial functions. These descriptors consider both the radial distribution and the angular distribution of the neighborhood surrounding each atom. By incorporating information from both the radial and angular aspects, the descriptors capture the local atomic environment in a more comprehensive manner, enabling a more accurate representation of the atomic energy within the MTP framework.
\begin{multline*}\tag{5} 
%begin{equation} \tag{2} 
M_{\mu,\nu}(r_{ij}, \tau_i, \tau_j) = \\
\sum_jf_\mu(|r_{ij}|,\tau_i, \tau_j)\underbrace{r_{ij}\otimes\cdots\cdots\cdots\otimes r_{ij}}_\text{{$\nu$ times}}
\end{multline*}
The radial function $f_{\mu,}$ is further expanded using radial basis functions $Q^{(\alpha)}$ and fitting parameter $c_{\mu,\tau_i,\tau_j}^{(\alpha)}$ as expansion coefficients. This expansion allows for a more flexible and accurate representation of the radial dependence of the atomic interactions.
\begin{equation} \tag{6}
f_{\mu,}(|r_{ij}|, \tau_i, \tau_j)=\sum_\alpha c_{\mu,\tau_i,\tau_j}^{(\alpha)}Q^{(\alpha)}(|r_{ij}|)
\end{equation}
The model parameters $\Theta = (c_{\beta}, c_{\mu,\tau_i,\tau_j})$ are determined during the minimization of the cost function as given by Equation \ref{eq:7}.  

\subsection*{Data curation and optimization}

We acquired \textit{ab initio} data using established methods and databases from prior research. 
The database construction involved two methodologies, namely manual processing \cite{bartok2018machine,zuo2020performance,zongo2022first} and active learning \cite{podryabinkin2017active} , as explained deeply in supporting information.  Specifically, we referred to the dataset created for the Gaussian approximation potential for silicon \cite{bartok2018machine}, the comparative study \cite{zuo2020performance}, and active learning techniques detailed in Ref. \cite{podryabinkin2017active}.
For liquid silica, we utilized the temperature range (1000 K - 5000 K) from previous databases specifically designed for neural network inter-atomic potentials (NNIP), which covered temperatures exceeding its boiling point and extended as high as 5000 K, as referenced in\cite{balyakin2020deep}. While NNIP potentials involve a very large number of adjustable parameters--typically tens of thousands--allowing to jointly describe a large number of off-equilibrium configurations, accommodating such deviations from equilibrium becomes challenging within the MTP framework due to limited numbers of parameters.  Effective MTP training, therefore, relies on carefully selecting the training set. Our final training dataset, herein referred to as the unified training set, was constructed through a two-step process: curation and subsequent optimization. 

To enhance the quality of our training set and to properly assess the error of the test set , down selection was applied. To begin, we sorted our full database into smaller subsets as elaborated in Tab. S3 through S6 in the supporting information. Then, within each subset, we utilize a filtering strategy referred to herein as the "train-remove-train" approach. We first train while monitoring for significant reductions in energy and force errors associated to each configuration as we incrementally raise the MTP level by one unit (We focus on levels 08 to 14 for deformation and defects, while levels 16 to 18 are used for disordered structures). Next, we analyze the error reduction between 2 or 3 consecutive levels. If a significant decrease is not observed, we then eliminate configurations based on factors such as: 

\begin{enumerate}[label=(\theenumi)]

\item Total energy: configurations with similar total energies yet differing atomic coordinates to other in the training set, as well as those with fluctuating total energies but nearly identical atomic positions to others in the training set are removed. These configurations originated from relaxation of molecules, single-point calculations of unrelaxed defects, manually constructed unrelaxed jump paths, and strained configurations where lattice parameters or vectors were strained without corresponding adjustments to atomic positions and embedded dimer. 
\item Contributions from smearing: this can be primarily attributed to the extensive use of high-temperature AIMD simulations. Silica configurations generated from AIMD simulations, featuring a number of atoms greater than 36, are excluded if they exhibit smearing contributions to the total energy greater than zero.
\item Minimum interatomic distances (this criteria complements the total energy criteria): if multiple configurations from the same batch exhibit similar minimum distances, some are removed. This technique was mostly used for oxygen molecules. For example, we check the inter-atomic distance in the batch of relaxed O$_3$ molecules. We also discard embedded dimer configurations by comparing their minimum inter-atomic distances with those of the AIMD configurations. If the minimum inter-atomic distances are equal, we choose AIMD over embedded dimer configuration. 
\item Polymorphism: These configurations are derived from deformations following thermal expansion. Most of these configurations have been accurately computed and were included in the training database. However, some polymorphs arising from displacive phase transformations and polymorphs sharing the same lattice system with identical coordination numbers were excluded. In the case of displacive phase transformation, we retain the parent crystal and exclude the child crystal. When dealing with two polymorphs that share the same lattice system and coordination number, we typically choose one of them. 
\end{enumerate}
After removing these undesired configurations, where we reinitiate the training process incrementally, following a pattern akin to the first stage. This “train-remove-train” process is iterated until we attain a high level of confidence in the cleanliness of the subset--\textit{i.e}. all configurations included in the subset are associated with training errors that decrease as the MTP level increases. In the subsequent curation phase, we examine the possibility of extracting an even smaller subset from each previously cleaned subset. \cite{podryabinkin2017active}. To achieve this, we use the "select-add" command embedded in the MTP code, as described in Ref. \cite{novikov2020mlip}. We applied the "select-add" command to every cleaned subset.

\subsection*{Training set} 
In implementing the unified potential, we selected a range of configurations from the optimized and curated database, as outlined in the Data Curation and Optimization section. Similarly, the test set was chosen from the same optimized database to eliminate any overlap between the two sets.
While the test set encompasses all properties or types of configurations represented in the extensive database, the training set only includes certain configuration types. This strategic approach aims to ensure the portability of the interaction potential.
Note that the database contains configurations of substantial size, with up to 1000 atoms. However, we constrained the maximum box size in the training set to 36, except for the interfaces set where a few configurations are of size 80. Consequently, the number of atoms in the training set's boxes ranges from 1 to 80.
Conversely, the test set contains configurations of the maximum size found in the database.
The specific types of configurations represented in the training set are detailed in Tab \Ref{tab:Final_T}.
By restricting the training set to specific configuration types, we aim to enhance the generalizability of our potential by simulating properties not present in the training set.
Within the current implementation of the MTP potential, we do not employ a validation set in the typical manner used to estimate overfitting or underfitting during the training of neural network models. Instead, we opt to utilize an offline test set for this purpose.
To gauge the potential's portability more comprehensively, we perform molecular dynamic simulations for properties that were not included in the training set.
\begin{table} 
\centering
\begin{tabular}{lcccc}
\multicolumn{5}{c}{ } \\ 
\hline
Cleaned and opitmized subset & & & Total  \\
 \hline
Vacancy Si & & & 50   \\
Single interstitial Si &  & & 63  \\
Stacking fault Si &  &  & 40 & \\
Isolated atom Si and O &  & & 2 \\
Deformation Si &  & & 99 & \\
Deformation SiO$_{2}$ &  & & 139  \\
Liquid Si &  & & 217 & \\
Liquid SiO$_{2}$ &  & & 353  \\
Oxygen molecules &  & & 21  \\
Interface Si/SiO2 &  & & 164  \\
Embedded dimer (Si, SiO2)  &  & & 2 \\
SiO$_{2}$ slab  &  & & 9 & \\
\hline
Unified training set &  &  & 1159  \\
\hline
%& \multicolumn{1}{l}{} & \multicolumn{1}{l}{} \\ 
\end{tabular}
\caption{\label{tab:Final_T} Final refined and optimized training dataset derived from extensive uncurated database. Please refer to the supporting information for specific details about the number of atoms per configuration.}
\end{table} 
\subsection*{Training and validation}
The cost function, as described by Equation \ref{eq:7}, was minimized using the Broyden–Fletcher–Goldfarb–Shanno (BFGS) algorithm, a quasi-Newton optimization method implemented in the MTP framework. Fundamentally, training the MTP model with atomic configurations entails finding the parameter set \{$\Theta$\}  by solving the minimization problem presented in Equation \ref{eq:7}. We trained the refined unified training sets as detailed in Tab. S7 (supporting information) and Tab. \ref{tab:Final_T}. First, MTP potentials with parameter sets ranging from 300 to 1600, corresponding to levels 18 to 26, were used to train set 1 (Tab. S7 and Fig. S1(a)) as part of preliminary works, including optimization, training mode, and testing. The preliminary works are also presented in supplemental information from Fig. S5 to Fig. S8. At this stage, two training modes were employed: vibration mode and structures weighting mode. Specifically, the potential resulting from the vibration mode was used as input for the structures weighting mode training. The training process iterated until a desirable level of accuracy was achieved. As outlined in the supporting information, we assessed the validation error using the resulting potentials. We employed two independent validation sets, denoted as Validation 1 and Validation 2, which atom distributions are shown in the supplementary Fig. S2 and Fig. S3, respectively. Validation 1 was randomly selected concurrently with the training set from the curated database. On the other hand, Validation 2 consisted of the AIMD cooling and equilibration trajectories at 300 K. These validation sets were utilized as part of preliminary work. For the final implementation, we utilized the level 28. Based on the preliminary work, we opted for the structure weighting mode. The two-step training mode, as applied to large-size configurations in set 1 (Tab. S7 and Fig. S1 a) ), was deemed unnecessary. Given that our final training set (Tab. \ref{tab:Final_T}) comprises small cell configurations, we exclusively employed the structure weighting mode.
. We utilized the resulting potential to conduct both static calculations and molecular dynamic simulations, with the outcomes presented in the main text.
\begin{multline*}\tag{7}\label{eq:7}
\sum_{i=1}^{n}\big[w_e(E^{mtp}(x^{(i)}, \Theta) - E^{qm}(x^{(i)}))^{2} + \\ w_f\sum_{j=1}^{N_a(x^{(i)})}\big|F_j^{mtp}(x^{(i)}, \Theta) - F_j^{qm}(x^{(i)})\big|^{2}\\
+ w_{\sigma}\big|\sigma^{mtp}(x^{(i)}, \Theta) - \sigma^{qm}(x^{(i)})\big|^{2}\big] \, \to\, min
\end{multline*}
Here, ${E^{qm},F^{qm}, \sigma^{qm}}$ denotes the values of energy, force, and stress computed by the quantum mechanical approach (DFT), while ${E^{mtp},F^{mtp}, \sigma^{mtp}}$ represents the corresponding values obtained from the MTP model. $w_e$, $w_f$, $w_{\sigma}$ are the relative weights indicating the importance of the energy, the force and stress in optimization procedure.

\subsection*{Static calculations} 
\noindent This section summarizes the mathematical procedure used to determine static properties presented in the article.
\newline
For the chemical component with a formula ${X_lY_mZ_n}$, we calculate the cohesive energy as follows:

\begin{equation} \tag{8}
E_{coh}=E_{X_lY_mZ_n} - (lE_{X} + mE_{Y} + nE_{Z}). 
\end{equation}
 Where $E_{X_lY_mZ_n}$ represents the energy of the supercell of the compound, while $E_{X}$, $E_{Y}$, and $E_{Z}$ correspond to the energies of the isolated atoms. The subscripts $l$, $m$, and $n$ indicate the number of X, Y, and Z atoms present in the building block $E_{X_lY_mZ_n}$ of the material. Due to variations in the number of atoms within the primitive cell of each polymorph compared to the standard structural configuration, we normalize the cohesive energy by dividing it by the number of atoms present in the regular phase.

\noindent Points defects formation properties such as vacancy formation energy ($E^f_{v}$ ) was calculated using this equations: 
\begin{equation} \tag{9}
E^f_{v}=E_{N_{0}-1} - \frac{N_{0}-1}{N_{0}}*E_{N_{0}}. 
\end{equation}
For interstitial formation energy $E^f_{i}$, we used:

\begin{equation} \tag{10}
E^f_{i}=E_{N_{0}+1} - \frac{N_{0}+1}{N_{0}}*E_{N_{0}}. 
\end{equation}
In equations (9) and (10), $N_{0}$ and $E_{N_{0}}$ correspond to the number of atoms and total energy of a perfect supercell.

Particularly, vacancies in SiO$_{2}$ polymorphs were estimated considering a neutral state. Thus, the formation energy in SiO$_{2}$ polymorphs was calculated using:
\begin{equation} \tag{11}
E_{f}=E_{vac} - E_{bulk} + \mu_{O}.  
\end{equation} 
In this equation, E$_{vac}$ and E$_{bulk}$ represent the energy of the supercell containing the oxygen vacancy and the energy of the bulk supercell, respectively. The chemical potential is defined as half of the energy of a dioxygen molecule ($\mu_{O}$ = 1/2*E$_{O_{2}}$).

The equilibrium bulk modulus which correspond to the curvature of the energy-volume curve at its minimum was derived from the second order elastic constants \cite{pandit2023first}.  We calculate elastic stiffness constant $C_{ij}$ using central finite difference formula. 

\begin{equation} \tag{12}
C_{ij}=\frac{P^{(+\varepsilon_{j})}_i-P^{(-\varepsilon_{j})}_i}{2*\varepsilon_{j}}. 
\end{equation} 
where $P^{(+\varepsilon_{j})}_i$ is the $i^{th}$ component of the stress tensor when the configuration is strained only by $j^{th}$ component ($\varepsilon_{j}$) of the strain vector ($\vec{\varepsilon}$). After applying directional or isotropic deformation, the atomic positions undergo relaxation while the overall box size remains fixed.
We compute the generalized stacking fault energy $(\gamma{(u)})$ by incrementally shifting the upper crystal half along the slip direction and assessing energy differences per unit area (A) of the fault plane.
\begin{equation} \tag{13}
\gamma{(u)}=\frac{E(u) - E_{o}}{A}.  
\end{equation} 
where, $E_{o}$ represents the energy of the perfect crystal, while E(u) denote the energy of the supercell with the fault vector u which is directly proportional to the Burgers vector (b). 
Surface energy is also calculated using the following expression:
\begin{equation} \tag{14}
\gamma=\frac{E_{slab} - NE_{bulk}}{2A}.  
\end{equation} 
In this context, A refers to the area of the slab, N represents the number of atoms in the slab, while E$_{slab}$ and E$_{bulk}$ denote the total energy of the slab and the bulk energy per atom, respectively.

\subsection*{Si - SiO$_{2}$ Interface construction}
\noindent Previous reports indicate that defects are commonly observed at the interface Si - SiO$_{2}$ due to the imperfect matching of the two materials. 
To avoid considerable lattice mismatch, we utilize specific techniques. First, we rotate the alpha-quartz structure to achieve a tetragonal configuration. Next, we duplicate both the silicon crystal and alpha-quartz structure, ensuring that the lattice dimensions perpendicular to the interface direction closely match. This approach enables us to apply a small strain (less than 2 \% ) to the lattice vectors before forming the interface. Technically, the lattice mismatch $\alpha$ can be defined as the relative difference in lattice parameters between two crystalline materials, often expressed as a percentage or in terms of the absolute difference in lattice constants along specific crystallographic directions:
\begin{equation} \tag{15}
\alpha=\frac{n*L_{1} - m*L_{2}}{n*L_{1} + m*L_{2}}.  
\end{equation} 
Lattice duplication factors are represented by integers n and m; $L_{1}$ and $L_{2}$ denote the lattice parameters of a given direction. In both cases, symmetric and asymmetric interfaces were constructed for both oxygen-terminated and silicon-terminated quartz slabs, incorporating Si (100), Si (110), and Si (111) slab orientations. Our objective is not solely to construct a flawless interface representation of a naturally occurring or real-world interface, but rather to explore the versatility of the potential.
We then estimate the interface energy using:
\begin{equation} \tag{16}
\gamma{}=\frac{E_{S} - ( n_{SiO_{2}}*E_{SiO_{2}} + m_{Si}*E_{Si} ) }{A}.  
\end{equation} 
Where A represents the area of the interface, n$_{SiO_{2}}$ and m$_{Si}$ represent the number of  formula units of $SiO_{2}$ and ${Si}$ in the interface system. $E_{S}$ is the energy of the supercell containing the interface. The terms E$_{SiO_{2}}$ and E$_{Si}$ correspond to the energy of silica and silicon per formula unit, respectively.
%\textcolor{red}{
Due to the impractical size of duplicated models for energy computation via density functional theory (DFT), smaller superlattices were also constructed involving Si (100) interfacing with $\alpha$-quartz, $\beta$-cristobalite, $\alpha$-cristobalite, $\beta$-tridymite, and amorphous silica. This facilitated comparison between results obtained using MTP potentials and those from DFT. Each simulation box contains two distinct interfaces. Initially, these small models were relaxed at 0 K using MTP potentials. Subsequently, MTP-driven molecular dynamics (MD) simulations were performed at temperatures of 300 K, 500 K, 800 K, and 1200 K for 100 ps, and configurations were selected from the trajectories. The energies of these selected configurations, as well as the relaxed configurations, were then computed using DFT-based single-point energy calculations.

\subsection*{Molecular dynamics simulation}
\noindent \textit{Ab initio} molecular dynamics (AIMD) simulation was carried out using Quantum Espresso using the parameters as described in the section "\textit{Ab initio} calculations details."
The integration timestep was set to 1 fs for Si and 2 fs for SiO$_2$. The ionic temperature during simulations was controlled using velocity rescaling.

Force-field MD simulations were performed using the Large-scale Atomic/Molecular Massively Parallel Simulator (LAMMPS) software package \cite{thompson2022lammps}. While it is impractical to perfectly replicate the MD settings in Quantum Espresso within LAMMPS, we aimed to make them as close as possible. To generate disordered structures, we employed the velocity rescaling thermostat to control the temperature during the simulations. The time integration used the same time steps as in the AIMD simulations. For studying point defects diffusion in Si and self-diffusion in SiO$_2$, the Nosee-Hoover thermostat \cite{evans1985nose} was employed. The latter simulations were performed using a time step of 1 fs, and the damping parameter was set to 100 fs. 
\section*{Data availability}
\noindent The Si-O-SiO$_2$ database and the unified potentials can be found at \url{https://gitlab.com/Kazongogit/MTPu}.
\section*{Code availability}
\noindent The main codes used for this work are Quantum ESPRESSO (version 6.8), LAMMPS (version 2022), PHONOPY, and MLIP-2 (version 2). They are available at \url{https://www.quantum-espresso.org}, \url{https://lammps.sandia.gov}, \url{https://phonopy.github.io/phonopy} and \url{https://mlip.skoltech.ru} respectively.  Further details can be found in the GitLab repository \url{https://gitlab.com/Kazongogit/MTPu}. Additionally, Python scripts were written for data manipulation and processing; most are available on the GitLab repository.
\bibliographystyle{unsrt}
\bibliography{ref1}

\section*{Acknowledgements}
\noindent We thank the Digital Research Alliance of Canada for generous allocation of compute resources. Financial support was provided by the Natural Sciences and Engineering Research Council of Canada (NSERC), the Nuclear Waste Management Organization (NWMO), the Association canadienne-française pour l'avancement des sciences (ACFAS), and the Canada Research Chair on Sustainable Multifunctional Construction Materials.
\section*{Author contributions}
\noindent L.K.B initiated and coordinated the work, overseeing the research project. K.Z was responsible for developing the reference database and fitting the unified potential. K.Z performed all static calculations and molecular dynamic simulations using DFT, MTP, and semi-classical potentials. All the authors, K.Z, H.S, L.K.B, and C.O.P, contributed to the writing of the paper. C.O.P and L.K.B  supervised KZ. C.O.P and L.K.B provided computing resources.
\section*{Competing interests}
\noindent The authors declare no conflict of interest.
\end{document}